\documentclass[12 pt]{article}
\usepackage[utf8]{inputenc}
\usepackage[T1]{fontenc}
\usepackage[margin=2cm]{geometry}
\usepackage{stmaryrd}
\usepackage{amsmath,amsfonts,amssymb,mathtools,mathrsfs,amsthm}
\usepackage{bbm}
\usepackage[ruled,vlined, linesnumbered, french, onelanguage]{algorithm2e}
\usepackage{graphicx} 
\usepackage{tikz}
\usepackage{tikz-cd}
\usetikzlibrary{calc,patterns,angles,quotes}
\usetikzlibrary{automata,arrows,positioning,calc}
\usepackage{subcaption}
\usepackage{graphicx}
\usepackage{adjustbox}
\usepackage[english]{babel}
\usepackage{pgfplots}
\pgfplotsset{width=10cm,compat=1.9}
\usepackage{hyperref}
\usepackage{fancyhdr}
\usepackage[capitalise]{cleveref}
\crefname{subsection}{subsection}{subsections}
\definecolor{tas_lim_green}{RGB}{0, 171, 70}

\newcommand{\TASLim}{
\begin{minipage}[t]{0.9\textwidth}
    \raggedright 
    \tiny
    \vspace{0pt}
    \textbf{Property.}
    This document is not to be reproduced, modified, adapted, published translated in any material form in whole    or in part nor disclosed
     to any third party without the prior written permission of Thales Alenia Space.\\
    © 2024 Thales Alenia Space all rights reserved.  \\
    
  \end{minipage}

}

\fancypagestyle{plain}{

  \fancyhead{}

  \fancyhead[L]{\thepage}

  \fancyhead[R]{\nouppercase{\leftmark}}

  \fancyfoot{}

   \fancyfoot[C]{\TASLim}
}

\newtheorem{theo}{Theorem}[section]
\newtheorem*{theo*}{Theorem}
\newtheorem{prop}[theo]{Proposition}
\newtheorem{corollaire}[theo]{Corollary}
\newtheorem{lemme}[theo]{Lemma}
\newtheorem{defprop}[theo]{Def/Prop}
\newtheorem{defi}[theo]{Definition}
\newtheorem{remark}[theo]{Remark}

\newcommand{\LO}{\text{\normalfont{LO}}^N(\mathbb{R}^2)}
\newcommand{\LB}{\text{\normalfont{L}}^N(\mathbb{R}^2)}
\newcommand{\LR}{\text{\normalfont{LR}}^N(\mathbb{R}^2)}
\newcommand{\Faisc}{\text{\normalfont{Faisc}}^N(\mathbb{R}^2)}
\newcommand{\FaiscP}{\text{\normalfont{FaiscP}}^N(\mathbb{R}^2)}
\newcommand{\FaiscS}{\text{\normalfont{Faisc}}_{\text{\normalfont{Sym}}}^N(\mathbb{R}^2)}
\newcommand{\Aplus}{(\mathcal{A}_L^+)_j}
\newcommand{\Aminus}{(\mathcal{A}_L^-)_j}
\newcommand{\Car}{\text{\normalfont{Car}}}
\newcommand{\CarFaisc}{\text{\normalfont{Car}}^{\text{\normalfont{F}}}}
\newcommand{\sign}{\text{\normalfont{sign}}}
\newcommand{\signSym}{\text{\normalfont{sign}}^{\text{\normalfont{S}}}}
\newcommand{\FI}{F^{\text{\normalfont{inv}}}}

\newcommand{\N}{\text{\normalfont{N}}}
\newcommand{\NGOFFA}{\text{\normalfont{N}}^{\text{\normalfont{GOFFA}}}}
\newcommand{\NR}{\text{\normalfont{N}}^{\text{\normalfont{R}}}}
\newcommand{\NFaisc}{\text{\normalfont{N}}^{\text{\normalfont{F}}}}
\newcommand{\NFaiscSym}{\text{\normalfont{N}}^{\text{\normalfont{S}}}}

\newcommand{\I}{\text{\normalfont{I}}}
\newcommand{\IR}{\text{\normalfont{I}}^{\text{\normalfont{R}}}}
\newcommand{\IFaisc}{\text{\normalfont{I}}^{\text{\normalfont{F}}}}
\newcommand{\IFaiscSym}{\text{\normalfont{I}}^{\text{\normalfont{S}}}}
\newcommand{\gammaR}{\gamma^{\text{\normalfont{R}}}}
\newcommand{\gammaFaisc}{\gamma^{\text{\normalfont{F}}}}

\newcommand{\sgn}{\text{\normalfont{sgn}}}
\newcommand{\obsc}{\text{\normalfont{obsc}}}
\newcommand{\ras}{\text{\normalfont{graz}}}
\newcommand{\ObscRas}{\text{\normalfont{ObscGraz}}}

\newcommand{\footremember}[2]{%
    \footnote{#2}
    \newcounter{#1}
    \setcounter{#1}{\value{footnote}}%
}
\newcommand{\footrecall}[1]{%
    \footnotemark[\value{#1}]%
} 
\title{Classification of  obscuration-free reflective polygonal light beams }

\author{%
\begin{tabular}{c}
Pierre Franck\footremember{tas}{Thales Alenia Space, Cannes, France}\footnote{Now at: LAAS-CNRS, Toulouse, France}%
\\
\small\url{pierre.franck@lass.fr}\\ \\
Audric Drogoul\footrecall{tas}  \\
\small\url{audric.drogoul@thalesaleniaspace.com}
\end{tabular}
}
\date{}
\begin{document}

\pagestyle{plain}

\maketitle
\begin{abstract}
    In this paper, we study the connected components of an obscuration-free planar polygonal light beam space modeling light propagation in optical systems composed of reflective surfaces and a focal plane. Through homotopy construction, we demonstrate that the connected components of this space are in bijection with the connected components of the reflective polygonal chains space, whose elements are the polygonal chains with their respective mirrors' orientations taken into account. In order to prove this, we introduce a topological invariant that provides an intelligible way for opticians to name homotopy-equivalent obscuration-free optical configurations thanks to previous work with polygonal chains. 
\end{abstract}



\section{Introduction}
One of the major challenges in optical design is the exploration of optical configurations and their classification.
Typically, a standard approach involves exploring these configurations using \emph{brute force} \cite{Zhu:21,Papa:21}.
This approach is limited not only by its algorithmic complexity but also by the uncertainty of whether all possible configurations have been explored.
Thus, classification emerges as a necessary step to understand the set of admissible systems.
It also allows for exhaustive and more refined sampling of the admissible space.
It is therefore important to understand which topological features discriminate between similar configurations and those that differ. Let us note that similar issues have been intensively tackled in robot motion planning, where the admissible set corresponding to the set of states that a robot can reach without being  in conflict with its environment nor itself 
 can be modeled as a semi-algebraic set \cite{CapcoEldin,prebet2023computing,basu2014algorithms}. In this context, the question is to know if a robot can move from a configuration to another, what can be formulated as the study of the connected components of the admissible space. Similarly, the problem we aim to study in this article consists of replacing the collision function used in robotics with a function of obscuration and reflection of a system, i.e., a mirror should not cut the light flux, and the light flux should not pass through the mirrors to which it is adjacent.

We consider a space $E$ of admissible optical configurations defined by  the absence of obscuration and the respect of the reflecting condition.
In this space $E$, all points within the same connected component correspond to optical systems sharing similar optical characteristics.
For two optical systems $X$ and $Y$, we aim to determine whether these two systems are connected by a continuous path within $E$.
This work can thus be seen as a preliminary step toward exploring the local minima of a cost function.
The continuous path would then correspond to the flow associated with the gradient vector field of the cost function, incorporating, for instance, optical and geometric criteria.

We propose to establish a topological nomenclature for optical systems and, ideally, to obtain an explicit classification that characterizes whether optical configurations $X$ and $Y$ belong to the same connected component.
To this end, we introduce the notion of a topological invariant.

\begin{defi}[Topological invariant] \label{def:invariant}
    A topological invariant is an application  $I: E \rightarrow B$
    that verifies for all $X, Y \in E$:
    \[
    \exists \; \gamma: [0,1] \overset{C^0}{\rightarrow} E \: \ \gamma(0) = X \text{ and } \gamma(1) = Y \implies \I(X) = \I(Y).
    \]
    The topological invariant is exact if the implication is an equivalence.
    \end{defi}
This notion of  topological invariant allows for an intelligible naming of the connected components of $E$.
Moreover, in cases where we do not have an explicit description of the components of $E$, the search for a topological invariant becomes a crucial step for their study  (their number and topological properties). For applications in optical design, this way of naming optical systems will be very useful for classifying solutions and helping the optician to effectively filter and optimize the entire set of admissible solutions obtained by brute force, which amounts to hundreds of thousands.

In this paper, we focus on the geometry of unfolded optical systems in the plane $\mathbb{R}^2$, and thus on the placement of mirrors within this space.
We examine the geometry formed by the telescope in the plane, viewing it as an object of Euclidean geometry.

In subsection \ref{subsection:Faisc}, we introduce the concept of a \emph{beam}, which is central to our study and serves as the geometric object used to model a light beam propagating within an optical system.
A \emph{beam} is a polygonal representation of an optical system: the surfaces are modeled as segments, their curvatures are neglected, and the adjacent light flux between two optical surfaces is modeled as the area enclosed by the two extreme rays.

\begin{figure}[h!]
    \centering
    \begin{subfigure}{0.5\textwidth}
    \includegraphics[width = \textwidth]{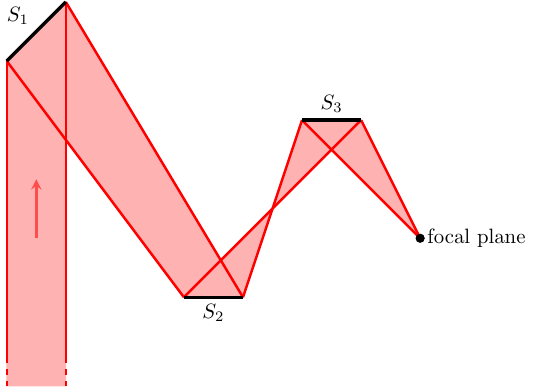}
    \caption{}
    \end{subfigure}
    \begin{subfigure}{0.45\textwidth}
    \includegraphics[width = \textwidth]{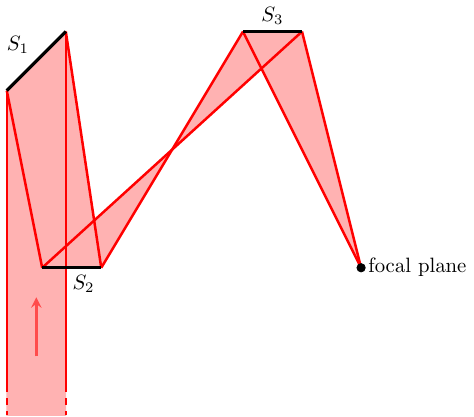}  
    \caption{}
    \end{subfigure}
    \caption{An obscuration-free (a) and and obscured (b) light beam composed of three mirrors and a focal plane. }
    \label{fig:faisceau_intro}
\end{figure}

 \Cref{fig:faisceau_intro}-(a) illustrates a beam $F$ corresponding to a telescope consisting of a source located at infinity, three mirrors, and a focal plane at a finite distance.
These three mirrors are depicted in the figure as $S_1$, $S_2$ and $S_3$.
The light flux follows this order, first arriving at $S_1$, then reflecting off $S_2$, subsequently reaching $S_3$, and finally converging onto the focal plane.
Note also that the mirrors and the focal plane do not obstruct the light flux; we say that the system is not \emph{obscured}.
When a surface obstructs the light flux, the beam is obscured and is thus not admissible (see \Cref{fig:faisceau_intro}-(b)).
This is one of the two essential constraints for an optical system to qualify as a beam.
The other constraint is intrinsic to the physics of mirrors, namely, reflection.
A beam is therefore a system comprising mirrors and a focal plane that satisfies the following two conditions: 
\begin{enumerate}
\item[(i)] Reflection or non-grazing incidence, stemming from the property of mirrors to reflect light.
\item[(ii)] Absence of obscuration, ensuring that the light flux is fully captured on the focal plane.
\end{enumerate}

\noindent
The following theorem is the main result of this paper.
\begin{theo*}[\ref{theoreme_nomenclature} and \ref{thm invariance}]
    The space of beams for $N = 3$ and $4$ mirrors has an exact topological invariant $\IFaisc$. 
\end{theo*}
This result tells us that two beams $F$ and $F'$ belong to the same connected component if and only if $\IFaisc(F)=\IFaisc(F')$.
It also provides the number of connected components of the space of beams for $3$ or $4$ mirrors.

To establish this theorem, we first model optical systems as polygonal chains, a simpler representation that is discussed in \cite{offaxis24}.
This simplification makes it possible to define an explicit invariant for $N=3$ or $4$ mirrors.
From this case, we introduce the notion of a reflexive polygonal chain, which bridges the gap between polygonal chains and beams.
We then examine the conditions for obscuration-free and non-grazing incidence when mirrors and the focal plane are free to move within a plane.
We consider optical systems with increasingly accurate representations as the section progresses.
\\
The interest in classifying heuristically obscuration-free optical solutions is not new. Indeed, let us mention the work of Stone et al. \cite{two_mirror_classif_second_orderStone:94} who addressed the problem in the case of a two mirror planar symmetric conic-based systems, satisfying an obscuration-free constraint and a set of first-order equations such as telecentricity and total magnification and canceling the second-order aberrations associated to distorsion. Another important work has been done in
\cite{2020Trumper} where 14 obscuration-free classes are heuristically presented,  missing only two classes compared to the mathematically proven classification presented in our previous paper \cite{offaxis24}. 
The purpose of this paper is to extend and complete the topological study done in \cite{offaxis24}  (in which the light was modeled as a polygonal chain),  to a more realistic  mathematical model in which the light is modeled as the area enclosed by two polygonal chains (corresponding to the extreme rays in the entry of the system).
\\
The paper is organized as follows.
In  \cref{section:LO}, we present the framework established in \cite{offaxis24}, where telescopes are modeled as the space of polygonal chains in $\mathbb{R}^2$ denoted as $\LO$ (see  \cref{def:LO}) for which an exact invariant $\N$ (see  \cref{def nomenclature}) is presented. We then provide increasingly complex representations of an optical system with reflexive polygonal chains in  \cref{section:LR} and beams in  \cref{section:Faisc} where the main result is established in \Cref{thm invariance}. Finally, in  \cref{subsection:FaiscSym}, we introduce a variation of the results established for beams by incorporating the action of a symmetry group on the beams and we illustrate the method for three mirrors in \cref{section:application}. 

\section{Obscuration-free polygonal chains} \label{section:LO}
This section presents the polygonal chains model derived in \cite{offaxis24}, which is the cornerstone of the beam light classification presented in the next sections.   
Here, we give the sketched ideas and refer the reader to \cite{offaxis24} for more details. 

Let us consider \(N+2\) points \((c_i)_{i = 0, \dots, N+1}\) in \(\mathbb{R}^2\)  to model the system: the half-line \((c_0, c_1]\) represents the direction of the source, the points \((c_j)_{j=1, \dots, N}\) correspond to the mirrors, and \(c_{N+1}\) represents the focal plane.  
Thus, we work within the space of \(N\)-polygonal chains \(\LB\), defined by:
\[
\LB:= \bigl(\mathbb{R}^2 \bigr)^{N+2}
\]
with the topology induced by \(\bigl(\mathbb{R}^2 \bigr)^{N+2}\).  
Let us first introduce the notion of a ray for a polygonal chain.

\begin{defi}[Ray for a polygonal chain] \label{def:rayon ligne brisee}
Let \(L\) be an \(N\)-polygonal chain given by \((c_i) \in \LB\). We denote the \(j\)-th ray for \(j \in \llbracket 0, N \rrbracket\) as  
\[
R_j = 
\begin{cases}
    (c_0, c_1] & \text{if } j = 0, \\
    [c_j, c_{j+1}] & \text{if } j \in \llbracket 1, N \rrbracket.
\end{cases}
\]

\end{defi}
On $\LB$, $R_0$ can be empty if $c_0=c_1$ and $R_j$ for $j\in \llbracket 1, N \rrbracket$ can be reduced to a point if $c_j=c_{j+1}$. Hence, we are going to add constraints so that these degenerate cases are considered as not acceptable. 
The   admissible set will be defined as the subset of $\LB$ with the constraints that neither  obscurating nor grazing rays exists such that there is no loss of information.
To achieve this let us introduce for $L = (c_i) \in \LB$ 
\begin{itemize}
    \item 
    the obscuration clause $\obsc$ :  
\begin{equation}
\label{rem:obscuration}
\obsc = \bigvee_{0 \leq j \leq N} \bigvee_{\substack{1 \leq k \leq N+1 \\ k \neq j,j+1}} \bigl( R_j \cap c_k \neq \varnothing \bigr) \lor (R_j=\varnothing),
\end{equation}
\item the grazing clause $\ras$ as 
\[
\ras:= \bigvee_{1 \leq i \leq N} c_{i}\in (c_{i-1},c_{i+1}).
\]
\end{itemize}

\begin{figure}[h!]
  \begin{subfigure}[b]{.32\textwidth}
    \centering
      \includegraphics[width=\textwidth]{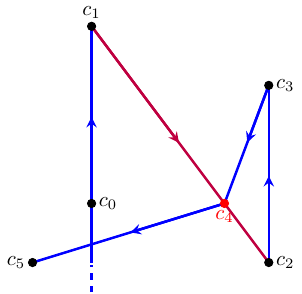}
      \caption{}
  \end{subfigure}
  \begin{subfigure}[b]{0.32\textwidth}
    \centering
       \includegraphics[width=\textwidth]{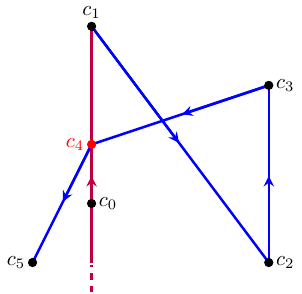}
      \caption{}
  \end{subfigure}
  \begin{subfigure}[b]{0.32\textwidth}
    \centering
       \includegraphics[width=\textwidth]{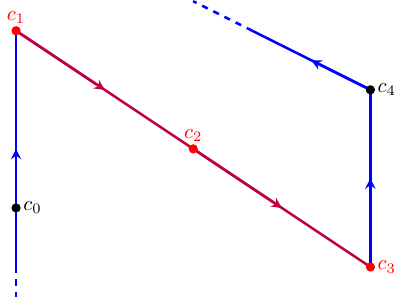}
      \caption{}
  \end{subfigure}
  \caption{Non admissible polygonal chains: (a) the mirror $c_4$ is obscuring the ray $R_1=[c_1,c_2]$, (b) the mirror $c_4$ is obscuring the light source $R_0=(c_0,c_1]$ and (c) the mirror $c_2$ is grazing since it belongs to $[c_1,c_3]$.}
  \label{fig:obscuration_grazing}
\end{figure}
In \cref{fig:obscuration_grazing}, we depict three $4$-polygonal chains  that are obscured. The one shown in (a) satisfies the condition $\obsc$ because $c_4$ belongs to the first ray $R_1=[c_1, c_2]$. The second one (depicted in (b)) satisfies the condition as well, since we observe that $c_4$ lies on the light source: $c_4$ belongs to $R_0=(c_0, c_1]$. Finally, the last one depicted in (c), satisfies $\ras$ because the points $c_1$, $c_2$, and $c_3$ are aligned, so that the system is grazing.

We introduce below the space studied for Off-Axis classification presented in \cite{offaxis24}. 

\begin{defi}[Space $\LO$] \label{def:LO}
We first define the joint clause of obscuration and grazing as $\ObscRas = \obsc \vee \ras$. The non-admissible polygonal chains subspace is defined as \( E_{\ObscRas} = \bigl\{ c \in \LB\,: \, \ObscRas(c) \bigr\} \). We then define the space of $N$-polygonal chains under obscuration and grazing constraints as:  
\[
\LO:= \LB \backslash E_{\ObscRas}.
\]
\end{defi}
Let us note that in \cite{offaxis24}, the clause conditions $\obsc$ and $\ras$ are expressed through polynomial inequalities, and thus $\LO$ is a semi-algebraic set. This fact raises the question about the number of connected components of $\LO$, as the number of connected components of a semi-algebraic set is bounded \cite[Theorem~5.22]{basubook}. Furthermore, it enables the use of relevant algorithms from real algebraic geometry, such as cylindrical algebraic decomposition, to identify at least one point in each connected component, which is the approach used by the authors.
\begin{defi} [Exact Invariant: Off-Axis Nomenclature] \label{def nomenclature}  
We introduce the nomenclature, denoted as  
\begin{align*}  
\N: \LO \rightarrow \pi_0 \bigl(\LO \bigr),  
\end{align*}  
where $\pi_0(\LO)$ is the set of connected components of $\LO$. We define $\N$ by taking $\N(L) = \N(L')$ if and only if $L$ and $L'$ are in the same connected component of $\LO$, for $L$ and $L'$ in $\LO$.  
\end{defi}
The nomenclature is therefore by definition an exact invariant in the sense of \cref{def:invariant}: it is even the canonical exact invariant since it directly uses the connected components of the space.
We have an explicit characterization of the nomenclature called the GOFFA (geometrical off-axis) nomenclature. This nomenclature, denoted by $\NGOFFA$, is introduced in \cite{offaxis24}, and the following theorem shows that it is an exact invariant in specific cases.

\begin{theo}[\cite{offaxis24}] \label{theoreme_nomenclature}
    The nomenclature given by $\NGOFFA: \LO \rightarrow B$ is an invariant of the space $\LO$. It is an exact invariant for $N= 3,4$ mirrors. Moreover, up to an  axial symmetry, $\text{LO}^3(\mathbb{R}^2)$ has  $16$ connected components and $\text{LO}^4(\mathbb{R}^2)$ has 144 connected components.
\end{theo}

Let us explain in more detail how the nomenclature $\NGOFFA$ assigns a name to each polygonal chain. The name of a system is given recursively on its mirrors. For a polygonal chain $L = (c_i)_{0 \leq i \leq N+1} \in \LO$, we process the mirrors step by step in the order $j = 1, \dots, N$.

At step $j$, we consider the segment $[c_j, c_{j+1}]$. If this segment does not intersect any of the previous segments $[c_k, c_{k+1}]$ for $1 \leq k < j$ or the light source $(c_0, c_1]$, then we assign the letter $V$ to the $j$-th mirror. Otherwise, we assign the letter $X$, and we also indicate as subscripts the segments and/or the half-line (for the light source) that are intersected by the segment $[c_j, c_{j+1}]$. 

Furthermore, we add the letter $A$ or $C$ ($A$ for \emph{anticlockwise} and $C$ for \emph{clockwise}) as a subscript to indicate the direction of rotation of the ray at the $j$-th mirror. This is explicitly given by the sign of $(c_{j+1}-c_j) \wedge (c_{j-1}- c_j)$, where $\wedge$ is the cross product operator. 
For $N=3$, these 3 invariants are sufficient. For $N=4$, an additional invariant that considers the set of points enclosed by a subset of the polygonal chain is necessary.
Thus, we observe that, up to symmetry with respect to the first mirror, all names begin with $V_A$.  

\section{Reflexive polygonal chains} \label{section:LR}
In this section, we introduce the space $\LR$ of reflexive polygonal chains, which provides a more accurate representation of optical systems than the space $\LO$ defined in the previous subsection. 
From now on, when there is no possibility of confusion, we denote $(c_i)$ as the sequence $(c_i)_{i\in I}$ when the index set $I \subset \mathbb{N}$ has been clearly specified beforehand.
 Mirrors are no longer just points; each now has an angle representing its direction. This introduces a new condition which is named \emph{reflection}. After defining this new space, we introduce two topological invariants:  
\begin{itemize}  
    \item[$\circ$] the nomenclature $\NR$, inspired by $\N$ defined in \cref{def nomenclature},  
    \item[$\circ$] the characteristic $\Car$ (see  \cref{def:caracteristique}).  
\end{itemize}  
Finally, we show that the joint invariant $\IR = (\NR \times \Car)$ is an exact invariant in \cref{prop_compo_LR}.
\begin{defi}[Reflexive polygonal chain] \label{def lignes brisees reflexives}
An $N$-reflexive polygonal chain is defined as a polygonal chain $(c_i)_{0 \leq i \leq N+1} \in \LO$ together with angles $(\theta_j)_{1 \leq j \leq N} \in (\mathbb{R}/2\pi \mathbb{Z})^N$ (associated with each mirror) that satisfy the following condition:  
\[
\forall \, \, 1 \leq j \leq N, \, \sgn \bigl((c_{j+1} - c_j) \cdot n(\theta_j) \bigr)=\sgn \bigl((c_{j-1} - c_j) \cdot n(\theta_j) \bigr) \neq 0
\]
where $n(\theta)$ is the unit normal of the line defined by the angle $\theta$, and $\sgn :\mathbb{R} \ni x \longrightarrow \sgn(x)\in\{ -1, 0, 1 \}$ is the sign function returning $1$ if $x>0$, $-1$ if $x<0$ and $0$ if $x=0$.
The space of reflexive polygonal chains is denoted as $\LR$.
\end{defi}

Geometrically, the condition for being a \emph{reflexive polygonal chain} means that $c_{j-1}$ and $c_{j+1}$ lie in the same half-plane bounded by the line passing through $c_j$ and  of angle $\theta_j$.

\begin{figure}[h!]
    \centering
    \begin{subfigure}[b]{.4\textwidth}
    \includegraphics[width=\textwidth]{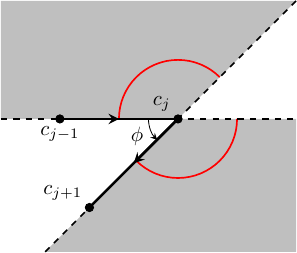}
    \end{subfigure}
    \begin{subfigure}[b]{.4\textwidth}
    \includegraphics[width=\textwidth]{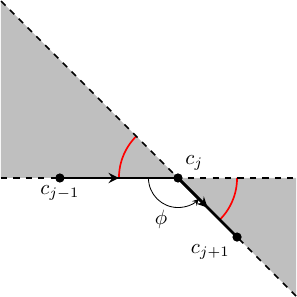}
    \end{subfigure}
    \caption{Admissible angles for reflexive polygonal chains.}
    \label{fig:reflexion}
\end{figure}

In \Cref{fig:reflexion}, we illustrate the points $c_{j-1}$, $c_j$, and $c_{j+1}$ for two different polygonal chains in $\LO$. The gray region in each case corresponds to the admissible area for the angle $\theta_j$ to satisfy the reflection condition at this point: the angle, represented as an element of the circle $\mathbb{R}/2\pi\mathbb{Z}$, must then belong to the red displayed circles. We observe that as the angle between the half-lines $[c_j, c_{j-1})$ and $[c_j, c_{j+1})$, denoted by $\phi$ in the figure, increases toward $\pi$, the space of admissible angles decreases. The limiting case where $\phi$ equals $\pi$ corresponds to the grazing case illustrated in  \Cref{fig:obscuration_grazing}-(c) for polygonal chains.

\begin{defi}[Application \(M_O\)] \label{def application M_O}
    An \(N\)-reflexive polygonal chain \(L\), given by \((c_i)\) and \((\theta_j)\), induces a polygonal chain in \(\LO\) via the application:
    \[
    \begin{aligned}
         M_O:\LR &\longrightarrow \LO \\
        L&\longmapsto (c_i)_{0 \leq i \leq N+1}.
    \end{aligned}
    \]
    where we recall that \(L = \bigl( (c_i), (\theta_j) \bigr)\).
\end{defi}

\begin{defi}[Ray of a Reflexive Polygonal Chain] \label{def:rayon_LR}
    Let \(L\) be an \(N\)-reflexive polygonal chain given by \(((c_i), (\theta_j))\). 
    The \(j\)-th ray is defined as the ray of \(M_O(L) \in \LO\) as defined in  \cref{def:rayon_LR}.
\end{defi}

\begin{defi}[Nomenclature \(\NR\)] \label{def:nomenclature reflexive}
    We define the nomenclature \(\NR\) for reflexive polygonal chains as the application:
    \[
    \begin{aligned}
        \NR: \LR &\longrightarrow \pi_0(\LO) \\
        L &\longmapsto \N((c_i)),
    \end{aligned}
    \]
    where \(L = \bigl( (c_i), (\theta_j) \bigr)\). 
    Thus, we have \(\NR = \N \circ M_O\).
\end{defi}

\begin{prop}[Topological Invariant: \(\NR\)] \label{prop invariance ligne brisee reflexive nomenclature}
    Let \(L\) and \(L'\) be in \(\LR\) such that there exists a continuous path \(\gammaR\) connecting \(L\) to \(L'\), then \(\NR(L) = \NR(L')\).
\end{prop}

\begin{proof}
    Straightforward by considering the application  \(M_O \circ \gammaR: [0,1] \rightarrow \LO\) which connects \(L_*:= M_O(L)\) to \(L_*':= M_O(L')\).
\end{proof}

It is also possible to define the reflection condition of reflexive polygonal chains using the angle function \( A: \mathbb{R}^2 \rightarrow \mathbb{R}/2\pi \mathbb{Z} \) which, for a non-zero vector \(x\), returns the angle \(\theta\) such that \( x = r \bigl( \cos( \theta), \sin(\theta) \bigr) \) for a unique \( r > 0 \); we will discuss this perspective further below.

\begin{defi}[Orientations of Mirrors] \label{def orientation miroirs}
    Let \( L \in \LO \) be an \( N \)-polygonal chain given by \( (c_i) \). We define the orientation spaces of the mirrors with the spaces \( (\Aplus \) (positive orientation) and \( (\Aminus \) (negative orientation) for the \( j \)-th mirror with \( 1 \leq j \leq N \) as:
    \begin{align*}
        \Aplus &= \bigl( A(c_j - c_{j-1}) + (-1)^j \, ]0, \pi [ \bigr) \cap \bigl( A(c_j - c_{j+1}) + (-1)^j \, ]0, \pi [ \bigr) \\
        \Aminus &= \bigl( A(c_j -c_{j-1}) + (-1)^{j+1} \, ]0, \pi [ \bigr) \cap \bigl( A(c_j - c_{j+1}) + (-1)^{j+1} \, ]0, \pi [ \bigr)
    \end{align*}
\end{defi}

Note that this also defines the orientation spaces of the mirrors for a reflexive polygonal chain \( L_R \in \LR \) by taking \( L = M_O(L_R) \).

\begin{remark} \label{convexite orientation miroirs}
    In  \cref{def orientation miroirs}, we defined the spaces \( \Aplus \) and \( \Aminus \) for a polygonal chain \( L \in \LO \) and \( 1 \leq j \leq N \). Note that these two spaces are convex as intersection of convex sets.
    
\end{remark}

We can then define reflexive polygonal chains from  \cref{def lignes brisees reflexives} equivalently using the mirror orientations. Indeed, given a polygonal chain $L = (c_i) \in \LO$ and angles $(\theta_j)\in(\mathbb{R}\backslash 2\pi\mathbb{Z})^N$, the data $((c_i),(\theta_j))$ satisfies the reflexive polygonal chain condition if and only if
\[
\forall \, 1 \leq j \leq N, \, \theta_j \in \Aplus \text{ or } \theta_j \in \Aminus.
\]

We will introduce a new topological invariant, in the sense of  \cref{def:invariant}, for the space $\LR$ called the \textit{characteristic}, which takes into account the positive or negative orientation of the mirrors (see  \cref{def orientation miroirs}).

\begin{defi}[Characteristic] \label{def:caracteristique}
Let $L$ be a $N$-reflexive polygonal chain given by the polygonal chain $(c_i)$ and angles $(\theta_j)$. The \textit{characteristic} of $L$, denoted $\Car(L)$, is an element of $\{ -1 ,1 \}^N$ and is defined as follows:
\[
\Car(L)_j = \begin{cases}
    1 \text{ if } \theta_j \in \Aplus, \\
    -1 \text{ if } \theta_j \in \Aminus,
\end{cases} \quad \text{for} \quad j \in \llbracket 1, N \rrbracket.
\]

\end{defi}

\begin{figure}[ht!]
    \centering
    \includegraphics[width=0.6\textwidth]{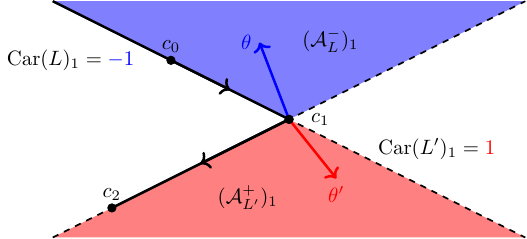}
    \caption{Caracteristic for two polygonal chains $L$ and $L'$}
    \label{fig: orientation}
\end{figure}

We represent the characteristic in  \Cref{fig: orientation} by depicting two 1-reflexive polygonal chains $L$ and $L'$. The first, denoted $L$, is given by $(c_0,c_1,c_2) \in \text{LO}^3(\mathbb{R}^2)$ and the angle $\theta \in \mathbb{R}/ 2\pi\mathbb{Z}$ as shown in the figure by the blue arrow. We then have $\Car(L)_1 = -1$ since $\theta$ is in the blue area representing $(\mathcal{A}_L^-)_1$. Similarly, the second reflexive polygonal chain $L'$ is also given by $(c_0,c_1,c_2)$ and the angle $\theta'$ represented on the figure by the red arrow. We then find $\Car(L')_1 = 1$ since $\theta'$ is in the red area representing $(\mathcal{A}_{L'}^+)_1$. Note that we have the following relations:
\begin{gather*}
    (\mathcal{A}_{L'}^-)_1 = (\mathcal{A}_L^-)_1, \quad
    (\mathcal{A}_{L'}^+)_1 = (\mathcal{A}_L^+)_1,
\end{gather*}
since $L$ and $L'$ share the same underlying polygonal chain, i.e., $M_O(L) = M_O(L')$.

\begin{prop}[Topological Invariant: $\Car$] \label{prop invariance ligne brisee reflexive car}
    Let $L$ and $L'$ be in $\LR$ such that there exists a continuous path $\gammaR: [0,1] \rightarrow \LR$ connecting $L$ to $L'$, then $ \Car(L) = \Car(L')$.
\end{prop}

\begin{proof}
    Suppose, by contradiction, that there exists a path $\gammaR: [0,1] \rightarrow \LR$ such that $\gammaR(0) = L$ and $\gammaR(1) = L'$ and some $k \in \llbracket 1, N \rrbracket$ such that 
    \[
    \Car(L)_k \neq \Car(L')_k.
    \]
    Without loss of generality, assume that $\Car(L)_k = 1$ and $\Car(L')_k = -1$. \\
    Let $\gammaR(t) = \bigl((c_i^t), (\theta_i^t)\bigr)$ where $(c_i^t) \in \LO$ and $(\theta_i^t) \in (\mathbb{R} / 2\pi \mathbb{Z})^N$. \\
    Specifically, we are interested in the evolution of the angle $\theta^t_k$ over time because it is at index $k$ that the characteristic of $L$ and $L'$ differs. To study this, we introduce the function 
    \begin{align*}
        \gamma_k^\theta: [0,1] &\longrightarrow \mathbb{R} / 2\pi \mathbb{Z} \\
        t &\longmapsto \theta_k^t - A(c_k^t - c_{k-1}^t)
    \end{align*}
    and it follows from the hypothesis that $R(\gamma_k^\theta(0)) \in (-1)^k \, ]0, \pi[$ and $R(\gamma_k^\theta(1)) \in (-1)^{k+1} ]0, \pi[$ (with $R$ being the bijection from $\mathbb{R} / 2\pi \mathbb{Z}$ to $ ]-\pi , \pi ]$), since $\theta_k^0 \in \mathcal{A}_k^+$ and $\theta_k^1 \in \mathcal{A}_k^-$. 
    Notice that the function $\gamma_k^\theta$ is continuous and the space $(\mathbb{R}/2\pi \mathbb{Z}) \backslash \{0 ,\pi \}$ decomposes into two connected components, $ ]-\pi, 0[$ and $ ]0, \pi[$. Thus, since $R(\gamma_k^\theta(0)) \in (-1)^k]0, \pi[$ and $R(\gamma_k^\theta(0)) \in (-1)^{k+1}]0, \pi[$, there exists some $t \in [0,1]$ such that $R(\gamma_k^\theta(t)) = 0$ or $R(\gamma_k^\theta(t)) = \pi$. 
    We then obtain $\sgn\bigl((c_{k-1}^t - c_k^t) \cdot n(\theta_k^t)\bigr) = 0$ because the line $(c_{k-1}^t, c_k^t)$ coincides with the line spanned by $\theta_k^t$ at $c_k$. This leads to a contradiction since the condition for a reflexive polygonal chain is violated.
\end{proof}

The following results will be useful to prove  \cref{prop_compo_LR}.

\begin{figure}
    \centering
    \includegraphics[width=0.6\textwidth]{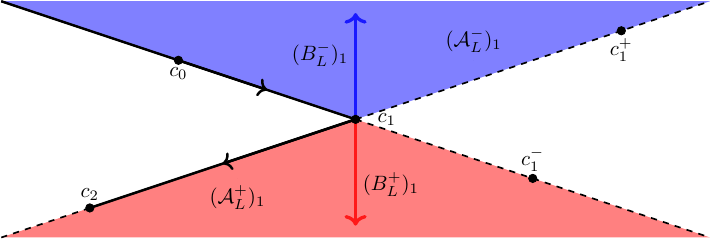}
    \caption{Illustration of $(B_L^+)_1$ and $(B_L^-)_1$}
    \label{fig: bissectrices off axis}
\end{figure}

\begin{defi}[Bisectors]
\label{def bissectors}
    Let $m_2$ be the covering map\footnote{We refer to Chapter 1 of \cite{Hatcher02}.
     }
     of $\mathbb{S}^1 :=  \mathbb{R}/ 2\pi \mathbb{Z}$ given by
    \begin{align*}
            m_2: \mathbb{R}/ 2\pi \mathbb{Z} &\longrightarrow \mathbb{R}/ 2\pi \mathbb{Z} \\
            x &\longmapsto 2x.
    \end{align*}
    Let $L \in \LO$ be given by $(c_i)$. We define the \textit{bisectors}  of $L$ as the elements in the fiber of $\psi(L)_j$ under $m_2$ (for $1 \leq j \leq N$), where
    \begin{align*}
        \psi: \LO &\longrightarrow (\mathbb{R} /2\pi \mathbb{Z})^N \\
        L=(c_i)_{0 \leq i \leq N+1} &\longmapsto \bigl( A(c_{j-1} - c_j) + A(c_j - c_{j+1}) \bigr)_{1 \leq j \leq N}
    \end{align*}
    and we denote by $(B_L)_j$ the fiber of $\psi(L)_j$ under $m_2$ which consists in two elements  \[
    (B_L)_j = \bigl\{ (B_L^+)_j, (B_L^-)_j \bigr\}
    \]
    where $(B_L^+)_j \in \Aplus$ and $(B_L^-)_j \in \Aminus$.
\end{defi}
\begin{remark}
    \label{re:bissectors}
    We can geometrically interpret $(B_L)_j$ by noting that it consists of the bisectors of the half-lines $[c_j, c_{j-1})$ and $[c_j, c^+_j)$ on one side, and $[c_j, c_{j+1})$ and $[ c_j, c^-_j )$ on the other side, where $c^\pm_j = c_j + (c_j - c_{j\pm1})$. We illustrate this in  \Cref{fig: bissectrices off axis} for $j = 1$. In this figure, we represent the orientations of the first mirror (at $c_1$) by the red part for $(\mathcal{A}_L^+)_1$ and the blue for $(\mathcal{A}_L^-)_1$. The red and blue arrows in the figure represent the bisection lines $(B_L^+)_1$ and $(B_L^-)_1$ of the reflexive polygonal chain, respectively. 
Thanks to this description, it is clear that $c_{j-1}$ and $c_{j+1}$ lie in the same half-plane defined by either of the bisection lines. 
\end{remark}


Before presenting the following lemma, which we will use to prove  \cref{prop_compo_LR}, we extend the definition of characteristic to work with arbitrary angles $\phi = (\varphi_j) \in (\mathbb{R}/ 2\pi \mathbb{Z})^N$ associated with an $N$-reflexive polygonal chain $L$, by defining
\[
\Car_L(\phi)_j = \begin{cases}
    1 \text{ if } \varphi_j \in \Aplus, \\
    -1 \text{ if } \varphi_j \in \Aminus, \\
    0 \text{ otherwise.}
\end{cases}
\]

\begin{lemme} \label{homotopie angles}
Let $L$ be an $N$-reflexive polygonal chain given by $(c_i) \in \LO$ and $(\theta_j) \in (\mathbb{R}/ 2\pi \mathbb{Z})^N$, and let $\phi = (\varphi_j) \in (\mathbb{R}/ 2\pi \mathbb{Z})^N$ such that $\Car_L(\phi) = \Car(L)$. Then $L$ and the $N$-reflexive polygonal chain $L'$, given by $(c_i)$ and $(\varphi_j)$, are in the same connected component of $\LR$.
\end{lemme}

\begin{proof}
    Using the notation from the lemma, we construct a path $\gamma: [0,1] \rightarrow \LR$ such that $\gamma(0) = L$ and $\gamma(1) = L'$ to show that $L$ and $L'$ are in the same connected component of $\LR$. 
    We define $\gamma$ as
    \begin{align*}
    \gamma: [0,1] &\longrightarrow \LR \\
    t &\longmapsto \bigl( (c_i), (\gamma_{\theta_j, \varphi_j}(t)) \bigr)    
    \end{align*}
    which is well-defined by convexity of the sets $(\mathcal{A}^{\pm}_L)_j$ (see \cref{convexite orientation miroirs}). 
\end{proof}

To show that by considering both topological invariants $\NR$ and $\Car$ together, we obtain an exact invariant $\IR:= (\NR \times \Car)$ of the space $\LR$, we prove the other implication of   \cref{def:invariant}.

\begin{prop} \label{prop_compo_LR}
Let $L$, $L'$ be elements of $\LR$ such that $\NR(L) = \NR(L')$ and $\Car(L) = \Car(L')$. Then there exists a path $\gamma_R: [0,1] \rightarrow \LR$ such that $\gamma_R(0) = L$ and $\gamma_R(1) = L'$. Therefore, $L$ and $L'$ are in the same connected component of the space $\LR$.
\end{prop}

\begin{proof}
    To prove the proposition, we first focus on $M_O(L)$ and $M_O(L')$, where we define $M_O: \LR \rightarrow \LO$ in \cref{def application M_O}. Since $\NR(L) = \NR(L')$, $M_O(L)$ and $M_O(L')$ are in the same connected component, so there exists a path 
    \begin{align*}
        P: [0,1] &\longrightarrow \LR
    \end{align*} 
    connecting $M_O(L)$ and $M_O(L')$.
    We now want to find a lift $\widetilde{P}_R: [0,1] \rightarrow \LR$ of $P$ through the map $M_O: \LR \rightarrow \LO$, which sends a reflexive polygonal chain $((c_i),(\theta_j))$ to $(c_i)$, such that $\widetilde{P}_R(0) = L_*$ and $\widetilde{P}_R(1) = L_*'$, where we define the reflexive polygonal chains $L_*$ and $L_*'$ in steps \ref{etape 1} and \ref{etape 2} below. These reflexive polygonal chains satisfy the following properties:
    \begin{itemize}
        \item[$\circ$] There exists a path $\gamma_L: [0,1] \rightarrow \LR$ (constructed in step \ref{etape 1}) such that $\gamma_L(0) = L$ and $\gamma_L(1) = L_*$,
        \item[$\circ$] There exists a path $\gamma_{L'}: [0,1] \rightarrow \LR$ (constructed in step \ref{etape 3}) such that $\gamma_{L'}(0) = L_*'$ and $\gamma_{L'}(1) = L'$.   
    \end{itemize}
    The lifting property of $\widetilde{P}_R$ is illustrated by the following commutative diagram:
    \begin{center}
        \begin{tikzcd} 
        {\{ 0 \}} \arrow[bend left = 15]{rrrr}{L_*} \arrow[hookrightarrow]{ddr} & {\{ 1 \}} \arrow[hookrightarrow]{dd} \arrow{rrr}{L_*'} &&& \LR \arrow{dd}{M_O} \\ \\ 
        & {[0,1]} \arrow{rrr}{P} \arrow[dashed]{rrruu}{\widetilde{P}_R} &&& \LO 
        \end{tikzcd}
    \end{center}

    To construct $\widetilde{P}_R$ (which is done in step \ref{etape 2}), we first observe that the first component of $\widetilde{P}_R$ is determined. Indeed, since $M_O \circ \widetilde{P}_R = P$ by commutativity of the diagram, we can write $\widetilde{P}_R (t) = (P(t), (\varphi^t_j))$ where the angles $\phi^t = (\varphi^t_j) \in (\mathbb{R}/ 2\pi \mathbb{Z})^N$ need to be determined. Additionally, we have
    \begin{gather*}
        M_O(L) = P(0) = M_O(L_*), \quad
        M_O(L') = P(1) = M_O(L_*').
    \end{gather*} 
    We proceed in three steps, which we detail below:
    \begin{enumerate}
        \item \label{etape 1} Starting from $L$ given by $(c_i) \in \LO$ and the angles $(\theta_j) \in (\mathbb{R}/ 2\pi \mathbb{Z})^N$, we want to construct a path from $L$ to $L_*$ given by $(c_i)$ and the angles $B^0 = (B^0_j)$ defined as $ B^0_j=(B_L^\pm)_j$ if $\theta_j\in (\mathcal{A}_L^\pm)_j$,
        with the notation from  \cref{def bissectors}. 
        Since $\Car_L(B^0) = \Car(L)$, there exists a path $\gamma_L: [0,1] \rightarrow \LR$ connecting $L$ to $L_*$ by  \cref{homotopie angles}.
        We now note that
        \[
        \Car(L) = \Car(L_*).
        \]
        \item \label{etape 2} Now, starting from $L_*$, we will construct a lift $\widetilde{g}_j$ of the path $g_j$ (for each $j$ in $\llbracket 1, N \rrbracket$), defined as
        \begin{align*}
            g_j: [0,1] &\longrightarrow \mathbb{R} / 2\pi \mathbb{Z} \\
            t &\longmapsto A\bigl( P_{j-1}(t) - P_j(t) \bigr) + A \bigl( P_j(t) - P_{j+1}(t) \bigr),
        \end{align*}
        by $m_2: \mathbb{R}/  2\pi \mathbb{Z} \rightarrow \mathbb{R}/  2\pi \mathbb{Z}$ such that $\widetilde{g}_j(0) = B^0_j$, as illustrated by the following commutative diagram:
        \[
        \adjustbox{scale=1.1,center}{
        \begin{tikzcd} 
            {\{ 0 \}} \arrow{r}{B^0_j} \arrow[hookrightarrow]{d} & \mathbb{R}/  2\pi \mathbb{Z} \arrow{d}{m_2} \\
            {[0,1]} \arrow{r}{g_j} \arrow[dashed]{ur}{\widetilde{g}_j} & \mathbb{R}/  2\pi \mathbb{Z}
        \end{tikzcd}
        }
        \]
        Since $m_2$ is a covering map, there exists a unique lift $\widetilde{g}_j$ of the path $g_j$ through $m_2$ such that $\widetilde{g}_j(0) = B^0_j$. Thus, denoting $\widetilde{g} = (\widetilde{g}_j)_{1 \leq j \leq N}$, we consider the continuous path
        \begin{align*}
            \widetilde{P}_R: [0,1] &\longrightarrow \LR \\
            t &\longmapsto (P(t), \widetilde{g}(t)),
        \end{align*}
        which is well-defined by  \cref{def bissectors}.
        This path satisfies $\widetilde{P}_R(0) = L_*$, and we define the endpoint of the path as $L_*':= \widetilde{P}_R(1)$. We now note that
        \[
        \Car(L_*) = \Car\bigl(L_*' \bigr)
        \]
        by  \cref{prop invariance ligne brisee reflexive car}.
        \item \label{etape 3} This step consists of reversing step \ref{etape 1}: we want to find a path $\gamma_{L'}$ from $L_*'$ to $L'$, with $L'$ given by $(c'_i) \in \LO$ and the angles $(\theta'_j) \in (\mathbb{R}/ 2 \pi \mathbb{Z})^N$. Note that $L_*' = \widetilde{P}_R (1) = (P(1), \widetilde{g}(1)) = ((c'_i), \widetilde{g}(1))$ because $P(1) = (c_i')$. Using the equalities
        \[
        \Car(L') = \Car(L) = \Car(L_*) = \Car \bigl( L_*' \bigr)
        \]
        by steps \ref{etape 1} and \ref{etape 2} and the hypothesis of the theorem, we obtain $\Car_{L'} (B^1) = \Car(L')$ where $B^1 = \widetilde{g}(1)$. Thus, there exists a path $\gamma_{L'}$ connecting $L_*'$ to $L'$ by  \cref{homotopie angles}.
    \end{enumerate}
    We must concatenate the paths given in steps \ref{etape 1}, \ref{etape 2}, and \ref{etape 3} in order to construct the path $\gamma_R$ that connects $L$ to $L'$. We then concatenate\footnote{Let two paths $\alpha, \beta: [0,1] \rightarrow X$ such that $\alpha(1) = \beta(0)$ with $X$ a topological space. We denote by $\alpha \cdot \beta$ the path
    \[
    t \mapsto \begin{cases}
        \alpha(2t) \text{ if } t \in [0,\frac{1}{2}], \\
        \beta(2t-1) \text{ if } t \in [\frac{1}{2},1].
    \end{cases}
    \]
    \label{footnote:concat}} the paths and take $\gamma_R:= (\gamma_L \cdot \widetilde{P}_R) \cdot \gamma_{L'}$. 
\end{proof}

\begin{corollaire}[Exact Invariant of $\LR$] \label{corollaire:LR invariant}
    We have a characterization of the connected components of the space of reflexive polygonal chains $\LR$ by the nomenclature $\NR$ and the characteristic $\Car$: we have
    \[
    \pi_0 \bigl(\LR \bigr) \simeq \pi_0 \bigl( \LO \bigr) \times \{ -1, 1 \}^N
    \]
    via the map 
    \begin{align*}
        \IR: \LR &\longrightarrow \pi_0 \bigl( \LO \bigr) \times \{ -1, 1 \}^N  \\
        L &\longmapsto \bigl( \NR(L), \Car(L) \bigr).
    \end{align*}
\end{corollaire}

\begin{proof}
Let $L$ and $L'$ be reflexive polygonal chains in $\LR$. If $L$ and $L'$ are in the same connected component, then we have  $\NR(L) = \NR(L') \, \text{ and } \,\Car(L) = \Car(L')$
since $\NR$ and $\Car$ are topological invariants by  \cref{prop invariance ligne brisee reflexive nomenclature} and \cref{prop invariance ligne brisee reflexive car}. 
Conversely, if we have $\NR(L) = \NR(L') \, \text{ and } \,\Car(L) = \Car(L')$
then $L$ and $L'$ are in the same connected component by  \cref{prop_compo_LR}. 
This shows that $\IR = (\NR \times \Car)$ is an exact topological invariant in the sense of  \cref{def:invariant}.

\medskip

We then have $\pi_0 \bigl(\LR \bigr) \cong \text{Im}(\IR)$. To show the result, we now prove that the map $\IR$ is surjective. Let $X_{\text{N}} \in \pi_0 \bigl( \LO \bigr)$ and $S \in \{ -1,1 \}^N$, then let $L \in \LO$ such that $\NR(L) = X_{\text{N}}$ and choose the angles $(\theta_j)$ with $\theta_j = (B_L^\pm)_j$ if $S_j=\pm1$,
for $j \in \llbracket 1, N \rrbracket$.
We then construct the reflexive polygonal chain $\widetilde{L} = \bigl(L, (\theta_j)\bigr)$ which satisfies $\IR \bigl(\widetilde{L} \bigr) = \bigl(X_{\text{N}}, S \bigr)$.
\end{proof}

\section{Polygonal beams} \label{section:Faisc}
In this section, we improve the space $\LR$ of reflexive polygonal chains by considering the more realistic space $\Faisc$ of beams. 
The section splits in two subsections:  \cref{subsection:Faisc} studies the space $\Faisc$ and  \cref{subsection:FaiscSym} considers $\FaiscS$  the quotient space by the symmetry group $\mathbb{Z}\backslash 2\mathbb{Z}$.
 
\subsection{General case} \label{subsection:Faisc}
In order to introduce the space $\Faisc$ of obscuration-free and non-grazing beams, we start by introducing the space of primary beams.

\begin{defi}[Primary beam]
    Let $N \in \mathbb{N}^*$, an $N$-primary beam is the data 
    \[
    \Bigl((a_i)_{0 \leq i \leq N+1}, (b_i)_{0 \leq i \leq N+1}\Bigr)\in \LB\times \LB
    \]
    of two polygonal chains of $\mathbb{R}^2$ verifying the following assumptions:
    \begin{itemize}
        \item[$\circ$] for $1 \leq k \leq N$, $a_k \neq b_k$,
        \item[$\circ$]  $a_0 = b_0=(-1,0)$ and $a_{N+1}=b_{N+1}$,
        \item[$\circ$] $ c_1=\frac{a_1 + b_1}{2}=(0,0)$.
    \end{itemize}
\end{defi}

In this definition, we exclude degenerate beams to ensure that the line $(a_0 c_1]$, representing the direction of the light source, is well-defined, and equal (without loss of generality)   to the half-axis of negative $y$-values and that the mirrors $[a_k,b_k]$ have a positive length. The condition $a_{N+1}=b_{N+1}$ means that the beam is convergent at the focal plane. The segment $S_i = [a_i,b_i]$ for $1\leq i\leq N$ denotes  the $i$-th mirror and $S_{N+1}=\{a_{N+1}\}$ is the focal plane.
We denote the space of $N$-primary beams $\FaiscP$. The topology we consider is the classical topolology induced by $\bigl(\mathbb{R}^2 \bigr)^{2(N+2)}$ on $\FaiscP$. 

\begin{defi}[The $t$-polygonal chain of a primary beam $F$] \label{def:t-ligne brisee}
Let $F$ be an $N$-primary beam  given by $(a_i)$ and $(b_i)$. 
Let $t \in[0,1]$, let us define $s_t$ that depicts the direction of the source for the $t$-polygonal chain of $F$ with $s_t = (1-t)a_1 + tb_1 + a_0$. \\
The $t$-polygonal chain $P_t$ is then defined recursively as the $N$-polygonal chain $P_t = (P_i^t)_{0 \leq i \leq N+1}$ with: 
\[
P_i^t = \begin{cases}
    s_t \text{ if } i = 0, \\
    (1-t)a_1 + tb_1 \text{ if } i = 1, \\
    \left.
    \begin{aligned}
    &(P_{i-1}^t, I_{i-1}^i) \cap [a_i,b_i] \text{ if } 
    (a_{i-1}, a_i) \cap (b_{i-1}, b_i) \neq \varnothing, \\
    &(1-\lambda_{i-1})a_i + \lambda_{i-1}b_i \text{ otherwise}.
    \end{aligned} \right\rbrace\text{ if }i\in\llbracket2,N+1\rrbracket
\end{cases}
\]
where $I_{i-1}^i = (a_{i-1}, a_i) \, \cap \, (b_{i-1}, b_i )$ when $(a_{i-1}, a_i) \, \cap \, (b_{i-1}, b_i) \neq \varnothing$ and $\lambda_k$ ($1 \leq k \leq N$) is the real number between $[0,1]$ such that $P_k^t = (1-\lambda_k)a_k + \lambda_k b_k $. \\
We say that $P_i^t$ is the $i$-th impact point of the $t$-polygonal chain.
\end{defi}

Let us notice that the extremal $t$-polygonal chains of $F$ are the polygonal chains $(a_i)$ and $(b_i)$, that is to say $P_0 = (a_i)$ and $P_1 = (b_i)$.
Thanks to  \cref{def:t-ligne brisee}, we have a parametrization of $t$-polygonal chains 
\begin{align*}
    P: [0,1] & \longrightarrow L^N(\mathbb{R}^2) \\
    t & \longmapsto P_t
\end{align*}
that connects the two polygonal chains $(a_i)$ and $(b_i)$. \\

\begin{defi}[Ray of a $t$-polygonal chain]  \label{def:rayon t-ligne brisee}
Let us consider the $t$-polygonal chain $P_t$ of the primary beam $F$ (see  \cref{def:t-ligne brisee}). We define for $j$ in $\llbracket 0, N \rrbracket$, the $j$-th ray $R_j^t$ of the $t$-polygonal chain as in  \cref{def:rayon ligne brisee}:  
\[
R_j^t = \begin{cases}
(P_0^t, P_1^t]  \text{ if } j=0, \\
[P_j^t, P_{j+1}^t] \text{ if }j\in \llbracket 1,N\rrbracket.
\end{cases}
\]
\end{defi}

\begin{defi}[Centered polygonal chain] \label{def:ligne brisee centrale}
Let $F$ be an $N$-primary beam given by the polygonal chains $(a_i)$ and $(b_i)$. We define the application $M$ as 
\begin{align*}
    M: \FaiscP &\longrightarrow \LB \\
    F &\longmapsto P_\frac{1}{2} = \Bigl( P_j^{\frac{1}{2}} \Bigr)_{0 \leq j \leq N+1}.
\end{align*}
We say that $M(F)$ is the centered polygonal chain of $F$. 
\end{defi}

\begin{defi}[Beam] \label{def:faisceau}
    Let $N \in \mathbb{N}^*$, an $N$-beam $F$ is an $N$-primary beam given by $(a_i)$ and $(b_i)$, verifying the following assumptions: 
        \begin{equation}
         \forall j\in \llbracket 0 , N\rrbracket\ \forall t\in [0,1]\  \forall k\in \llbracket 1 ,N+1\rrbracket \backslash\{j,j+1\}\ 
        R_j^t \cap S_k = \varnothing
        \tag{Non-obscuration}
        \label{condition non-obscuration}
        \end{equation}
        \begin{equation}
        \begin{aligned}
        \forall i\in \llbracket 1 , N\rrbracket\quad 
        \sgn \bigl((a_{i-1}-a_{i})\cdot n_i \bigr) = \sgn \bigl((b_{i-1}-b_{i})\cdot n_i \bigr) = \\
        \sgn \bigl((a_{i+1}-a_{i})\cdot n_i \bigr) =\sgn \bigl((b_{i+1}-b_{i})\cdot n_i \bigr) \neq 0
        \end{aligned}
        \tag{Reflexion}
        \label{condition reflexion} 
        \end{equation}
        where $n_i$ is the unit normal vector of $S_i$ and $\sgn$ the sign function.
\end{defi}

We denote $\Faisc$ the space of $N$-beams. The topology we consider is the topology induced by $\bigl(\mathbb{R}^2\bigr)^{2(N+2)}$. 
In  \Cref{fig:faisceau_intro}-(a),  \cref{def:faisceau} is depicted with the example of a $3$-beam. Let us notice that it indeed respects the conditions of non-obscuration and reflexion.


\begin{remark} [Reflexion] \label{remarque reflexion}
    The reflexion condition of a beam is equivalent to tell that the points $a_{i-1}$, $b_{i-1} $, $a_{i+1} $ and  $b_{i+1} $ are in the same half-plane of boundary $(a_i,b_i)$. Then, the convex hull $\text{\normalfont{Conv}}(a_{i-1}, b_{i-1}, a_{i+1}, b_{i+1})$ is in the same half-plane of boundary $(a_i,b_i)$.
\end{remark}
\begin{remark}\label{remarque semi-algebraic}
    We can show that the space of beams $\Faisc$ is a semi-algebraic set
    since 
   it expresses using only Euclidian geometry conditions which can be translated into polynomial inequalities.
\end{remark}

\begin{lemme}[Property of $t$-polygonal chains of $F$] \label{lemme:t-ligne brisee}
    Let $F \in \Faisc$, every $t$-polygonal chain $P_t$ of $F$ (see \cref{def:t-ligne brisee}) belongs to $\LO$ for $t$ in $[0,1]$.
\end{lemme}

\begin{proof}
    We will show the contraposition and we consider then that $P_t$ (for $t$ in $[0,1]$) is a polygonal chain verifying $\ObscRas$ of \cref{def:LO}. We then have two possibilities:
    \begin{itemize}
        \item[$\circ$] $L$ verifies $\obsc$. There exists $0 \leq j \leq N$ and $1 \leq k \neq j, j+1 \leq N+1$ such that $P_j^t \cap c_k \neq \varnothing$ by  \eqref{rem:obscuration}. Thus, we have $R_j^t \cap S_k \neq \varnothing$ as $c_k \in S_k$. Hence, the non-obscuration condition is not verified and $F$ is not a beam then.
        \item[$\circ$] $L$ verifies $\ras$. There exists $1 \leq i \leq N$ such that $P_i^t$ belongs to $[P_{i-1}^t,P_{i+1}^t]$. Therefore, we have $[P_{i-1}^t,P_{i+1}] \cap (a_i b_i) \neq \varnothing$ as $P_i^t$ belongs to $[a_i,b_i]$. Thus, $\text{Conv}(a_{i-1},b_{i-1},a_{i+1},b_{i+1})$ is not in the same half-plane of boundary $(a_i b_i)$, the reflexion condition is then not verified (\cref{remarque reflexion}).
    \end{itemize}
\end{proof}

\begin{remark} \label{rem:ligne brisees centrale faisceau}
    With  \cref{lemme:t-ligne brisee}, the application $M$ defined in \cref{def:ligne brisee centrale} induces an application $\widetilde{M}: \Faisc \rightarrow \LO$ where $\widetilde{M} = M|_{\Faisc}$. We denote this application $M$ in the sequel by abuse of notation. 
\end{remark}

\begin{defprop}[Application $M_R$]
    An $N$-beam $F$ given by $(a_i)$ and $(b_i)$ induces a reflexive polygonal chain through the application:
    \begin{align*}
        M_R: \Faisc & \xlongrightarrow{(M,A)} \LR \\
        F & \longmapsto \Bigl(M(F), \bigl(A(a_j-b_j)\bigr)_{1 \leq j \leq N} \Bigr)
    \end{align*}
    where $M: \Faisc \rightarrow \LO$ is given by  \cref{rem:ligne brisees centrale faisceau} and $A: \mathbb{R}^2 \rightarrow \mathbb{R}/2 \pi \mathbb{Z}$ is the angle function that returns the angle $\theta$ to $x \neq 0$ such that $x = r \bigl(\cos( \theta), \sin(\theta) \bigr)$ for a unique $r > 0$.
\end{defprop}
    
\begin{proof}
    We have that $M(F)$ belongs to $\LO$ by  \cref{lemme:t-ligne brisee}. Moreover, as the beam $F$ verifies the reflexion condition of beams, we have that $a_{j-1}$, $b_{j-1} $, $a_{j+1} $ and $b_{j+1} $ are in the same half-plane of boundary $(a_j,b_j)$, by  \cref{remarque reflexion}, for $j$ in $\llbracket 1, N \rrbracket$. This implies that points $c_{j-1}$ and $c_{j+1} $ are in the same half-plane with the boundary  line passing through $c_j$ with angle $\theta_j$ where we consider $(c_i):= M(F)$ and $(\theta_j):= (A(a_j - b_j)) $.
\end{proof}

\begin{defi} [Beam homotopy]
Let us consider two $N$-beams $F$ and $F'$ given respectively by $\bigl((a_i),(b_i)\bigr)$ and $\bigl((a_i'),(b_i')\bigr)$, we say that $F$ and $F'$ are \textit{homotopic} if there exists an application 
\[
H: \bigl( \llbracket 0, N+1 \rrbracket \sqcup \llbracket 0, N+1 \rrbracket \bigr) \times [0,1] \longrightarrow \mathbb{R}^2
\]
such that $\begin{cases}
H(\cdot, 0) = F \text{, that is to say } H(p_1(i), 0) = a_i \text{ and } H(p_2(i), 0) = b_i \\
H(\cdot, 1) = F' \\
H(\cdot, t) \in \Faisc \text{ for all } t \text{ in } [0,1]
\end{cases}$ \\
where $p_1, p_2: \llbracket 0, N+1 \rrbracket \longmapsto \llbracket 0, N+1 \rrbracket \, \sqcup \, \llbracket 0, N+1 \rrbracket$ are the two canonical applications of the union. 
We denote $F \cong F'$ when $F$ and $F'$ are homotopic.
\end{defi}

Let us notice that $F \cong F'$ is equivalent to the existence of a continuous path $\gamma: [0,1] \rightarrow \Faisc$ such that $\gamma(0) = F$ and $\gamma(1) = F'$.

Before giving the next proposition, we recall that a polygonal chain $(z_i)_{0 \leq i \leq N+1}$ in $\LO$ has a nomenclature $\N \bigl((z_i)\bigr)$ which returns the connected component of $\LO$ in which $(z_i)$ belongs (see  \cref{def nomenclature} for more details).

\begin{prop} \label{nomenclature faisceau}
Let $F$ be an $N$-beam given by $(a_i)$ and $(b_i)$, then $\N \bigl((a_i)\bigr) = \N \bigl((b_i)\bigr)$, that is $(a_i)$ and $(b_i)$ are in the same connected component of $\LO$. \\
More generally, we have for $t, t'$ in $[0,1]$, $\N(P_t) = \N(P_{t'})$. 
\end{prop}

\begin{proof}
    We use the parametrization of rays $\widetilde{P}:t\ni[0,1]\mapsto P_t\in \LO$ which is well defined by  \cref{lemme:t-ligne brisee}.
    Moreover, we have $\widetilde{P}(0) = (a_i)$ and $\widetilde{P}(1) = (b_i)$. So $\widetilde{P}$ is a path in $\LO$ connecting  $(a_i)$ and $(b_i)$, they are then in the same connected component and we obtain $\N\bigl((a_i)\bigr) = \N\bigl((b_i)\bigr)$. 
    We also use the application $\widetilde{P}$ to show that $\N(P_t) = \N(P_{t'})$ with $t,t' \in [0,1]$ by remarking that $\widetilde{P}(t) = P_t$ and $\widetilde{P}(t') = P_{t'}$; we then have a path between $P_t$ and $P_{t'}$ in $\LO$.
\end{proof}

\begin{defi}[$\NFaisc$ nomenclature]
    We define the nomenclature $\NFaisc$ of the space of beams $\Faisc$ as
    \begin{align*}
        \NFaisc: \Faisc &\longrightarrow \pi_0(\LO) \\
        F &\longmapsto \N (M(F))
    \end{align*}
    where $M$ is the application defined in  \cref{rem:ligne brisees centrale faisceau}.
\end{defi}

 \Cref{nomenclature faisceau} tells us that the nomenclature of a beam does not depend on the chosen ray: we could as well take $\N \bigl((a_i)\bigr)$ or $\N \bigl((b_i)\bigr)$ instead of $\N \bigl(M(F)\bigr)$ since these nomenclatures are all the same. Thus, the $\NFaisc$ nomenclature is a well-defined notion for a beam.

\begin{defi}[Characteristic of a beam: $\CarFaisc$]
    Let $F$ be an $N$-beam. The \textit{characteristic} of the beam $F$ is defined by extension as  
    \[
    \CarFaisc(F):= \Car(M_R(F)).
    \]
\end{defi}

\begin{figure}[h!]
    \centering
    \label{fig:a}
    \begin{subfigure}{0.45\textwidth}
         \includegraphics[width=\textwidth]{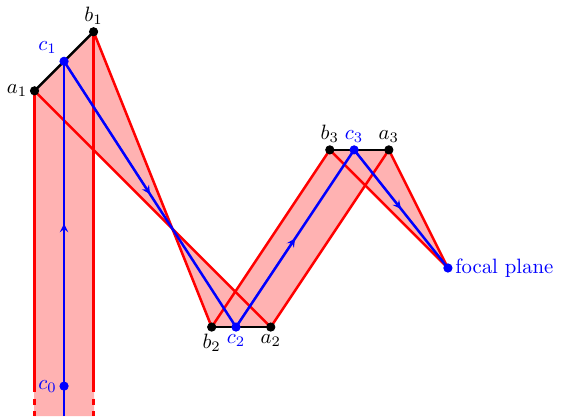}
    \end{subfigure}
    \begin{subfigure}{0.45\textwidth}
         \includegraphics[width=\textwidth]{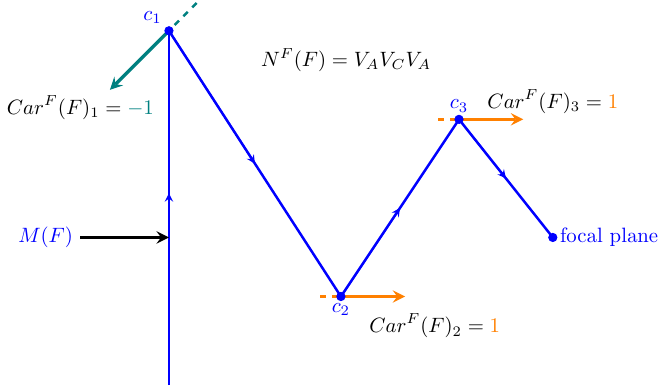}
    \end{subfigure}
    \caption{A beam $F$, its nomenclature and characteristic}
    \label{fig:faisceau invariant}
\end{figure}

In \Cref{fig:faisceau invariant}, we represent on the left the beam we had shown in \Cref{fig:faisceau_intro}-(a) with the centered polygonal chain $M(F)$ added in blue in the figure. On the right, we then consider $M(F)$ only and we give the nomenclature of the beam as we represent the angles $\theta_j$ of $M_R(F)$ with arrows (in orange when $\theta_j \in \Aplus$ and in green when $\theta_j \in \Aminus$).

So far, we have introduced the $\NFaisc$ nomenclature and $\CarFaisc$ characteristic. We use these two in the application $\IFaisc$ we now define in the following proposition.

\begin{prop}[Topological invariant: $\IFaisc$] \label{prop:invariance}
    Let $F$ and $F'$ be two beams in $\Faisc$ such that $F \cong F'$, then
    $
    \IFaisc(F) = \IFaisc(F')
    $
    where we define the application $\IFaisc$ as 
    \begin{align*}
        \IFaisc:= \bigl( \NFaisc \times \CarFaisc \bigr): \Faisc &\longrightarrow \pi_0(\LO) \times \{ -1, 1\}^N \\
        F &\longmapsto \bigl(\NFaisc(F), \CarFaisc(F)\bigr).
    \end{align*}
\end{prop}

\begin{proof}
    Let us consider $F$ and $F'$ such that $F \cong F'$, then, there exists  a path $\gammaFaisc$ in $\Faisc$ between $F$ and $F'$. Consider now the function $M_R \circ \gammaFaisc: [0,1] \rightarrow \LR $ that connects $L:= M_R(F)$ to $L':= M_R(F')$, we have 
    $
    \IR(L) = \IR(L')
    $
    since $\IR:= \bigl( \NR \times \Car \bigr)$ is a topological invariant by  \cref{corollaire:LR invariant}. This means that we have
    $\NR(L) = \NR(L')$ and $\Car(L) = \Car(L')$,
    and we then find    
        $\NFaisc(F) = \NR(L) = \NR(L') = \NFaisc(F')$ and
        $\CarFaisc(F) = \Car(L) = \Car(L') = \CarFaisc(F')$
    and eventually $\IFaisc(F)=\IFaisc(F')$.
\end{proof}

 In the following part, we introduce definitions and results that we use in order to show  \cref{thm invariance} which shows that $\IFaisc$ is an exact invariant.

\begin{defi}[$P_l$ application] \label{def application P_l}
We define the $P_l$ application with parameter\\ $l:= \bigl( (l_j^a)_{1 \leq j \leq N}, (l_j^b)_{1 \leq j \leq N} \bigr) \in \bigl((\mathbb{R}^+_*)^N\bigr)^2$ as 
\begin{align*}
    P_l: \LR &\longrightarrow \FaiscP \\
    \bigl((c_i), (\theta_j)\bigr) &\longmapsto \bigl((a_i),(b_i)\bigr)
\end{align*}
with 
\[
(a_i, b_i) = \begin{cases}
    (c_i,c_i) \text{ if } i \in \{0, N+1\} \\
    \Bigl( c_i + l_i^a \bigl(\cos(\theta_i),\sin(\theta_i)\bigr), c_i - l_i^b \bigl( \cos(\theta_i),\sin(\theta_i) \bigr) \Bigr) \text{ if } i \in \llbracket 1, N \rrbracket
\end{cases}
\]
\end{defi}

\begin{defi}[$d_R$ function]
    We introduce the continuous function $d_R$ representing the minimal mirror-to-mirror distance for a reflexive polygonal chain
    \begin{align*}
        d_R: \LR &\longrightarrow \mathbb{R}^+_* \\
        \bigl((c_i), (\theta_j)\bigr) &\longmapsto \min\limits_{1 \leq i \leq N} \min \Bigl(\bigl|(c_{i-1}- c_i) \cdot n(\theta_i)\bigr|, \bigl|(c_{i+1} - c_i) \cdot n(\theta_i)\bigr| \Bigr) 
    \end{align*}
    where $n$ is the function that gives the unit normal vector to an angle.
\end{defi}

The terms $|(c_{i-1}-c_i) \cdot n(\theta_i)|$ that we use in order to define the application, represents the distance between the point $c_{i-1}$ and the line $d(\theta_i)$ of direction $\theta_i$ and crossing $c_i$. In the same manner, $|(c_{i+1}-c_i) \cdot n(\theta_i)|$ is the distance between the point $c_{i+1}$ and the line $d(\theta_i)$.

\begin{prop} \label{prop faisceau P_l condition reflexion}
    Let $L$ be an $N$-reflexive polygonal chain given by $\bigl((c_i), (\theta_j)\bigr)$ and $P_l(L)$ the primary beam given by $\bigl((a_i),(b_i)\bigr)$  as stated in \cref{def application P_l}. Assume 
    \[
    K:= \max\limits_{1 \leq j \leq N} \max(l_j^a, l_j^b)  < d_R(L),
    \]
    then the reflexion condition of a beam (stated in \cref{condition reflexion}) is verified by $F$.
\end{prop}

\begin{proof}
First of all, we have by definition of $d_R$
\begin{gather*}
    d \bigl( c_{i-1}, (a_i, b_i) \bigr) \geq d_R(L) \; \text{ and } \; d \bigl( c_{i+1}, (a_i, b_i) \bigr) \geq d_R(L)
\end{gather*}
for all $i$ in $\llbracket 1, N \rrbracket$, where for $x \in \mathbb{R}^2$ and $Y \subseteq \mathbb{R}^2$, we define $d(x, Y)$ as 
\[
d(x, Y):= \inf\limits_{y \in Y} |x-y|
\]
We indeed observe that $
    d \bigl( c_{i-1}, (a_i b_i) \bigr) = |(c_{i-1} - c_i) \cdot n(\theta_i)|$ and $
    d \bigl( c_{i+1}, (a_i b_i) \bigr) = |(c_{i+1} - c_i) \cdot n(\theta_i)|$.
By  \cref{remarque reflexion}, the reflexion condition returns to verifying for some $N$-beam given by $\bigl( (x_i),(y_i) \bigr)$ if the points $x_{i-1}$, $y_{i-1}$, $x_{i+1}$, $y_{i+1}$ are in the same half-plane of boundary $(x_i ,y_i)$ for $i$ in $\llbracket 1, N \rrbracket$. In our case, we have $c_{i-1}$ and $c_{i+1}$ in the same half-plane of boundary $d(\theta_i)$, the line produced by the angle $\theta_i$ as $L$ is in $\LR$.
If we then take $K < d_R(L)$,
we have $|a_{i-1} - c_{i-1}| = l_{i-1}^a < d_R(L)$ and $d\bigl(c_{i-1}, (a_i, b_i) \bigr)  \geq d_R(L)$ for all $i$ in $\llbracket 1,N \rrbracket$ and so $a_{i-1}$ and $c_{i-1}$ are in the same half-plane of boundary $(a_i,b_i)$. We can do the same reasoning with the couples of points $(b_{i-1}, c_{i-1} )$, $(a_{i+1}, c_{i+1})$ and $(b_{i+1}, c_{i+1})$ to show that they are in the same half-plane of boundary $(a_i ,b_i)$. 
Thus, the points $a_{i-1}$, $b_{i-1}$, $a_{i+1}$ and $b_{i+1}$ are in the same half-plane of boundary $(a_i,b_i)$ and the reflexion condition is then verified.    
\end{proof}

\begin{defi}[$d_O$ function]
    We introduce the continuous function $d_O$ representing the distance of the rays to the mirrors (not adjacent) for a reflexive polygonal chain
    \begin{align*}
    d_O: \LR &\longrightarrow \mathbb{R}^+_* \\
    \bigl((c_i), (\theta_j)\bigr) &\longmapsto \min_{0 \leq j \leq N} \min_{\substack{1 \leq k \leq N+1 \\ k \neq j,j+1}} d(c_k,R_j)
    \end{align*}
    where $d(x,Y) = \inf\limits_{y \in Y} |x-y|$ with $x \in \mathbb{R}^2$ and $Y \subseteq \mathbb{R}^2$ and where $R_j$ is the $j$-th ray ray of the reflexive polygonal chain (see  \cref{def:rayon_LR}).
\end{defi}

\begin{lemme} \label{lemme faisceau P_l condition non obscuration}
    Let $L$ be an $N$-reflexive polygonal chain given by $\bigl((c_i), (\theta_j)\bigr)$ and $F:= P_l(L)$, with $l = \bigl( (l_j^a, l_j^b) \bigr)$, the primary beam given by $\bigl((a_i),(b_i)\bigr)$. We have the following results: 
    \begin{enumerate}
        \item \(
        \forall \, k \in \llbracket 1,N \rrbracket, \forall \, w \in S_k, |w-c_k| \leq \max(l_k^a, l_k^b)
        \) where $S_k = [a_k,b_k]$,
        \item 
        \(\forall \, j \in \llbracket 0, N \rrbracket, \forall \, t \in [0,1], \forall \, z \in R_j^t \text{ ($t$-polygonal chain of $F$)} \)
        \[
        \exists \, x \in R_j \text{ ($j$-th ray of $L$)}, \text{ such that } |z-x| \leq K
        \]
        with $K:= \max\limits_{1 \leq j \leq N} \max (l_j^a, l_j^b)$.
        
    \end{enumerate}
\end{lemme}

\begin{proof}
Result 1. comes directly from the definition of $P_l$. \\ 
We show result 2. by considering the set $
X_j = \bigcup\limits_{x \in R_j} B^f(x, K)$
with $j$ fixed and where $B^f(x, K)$ is the closed ball of center $x$ and of radius $K$.
We must show that all $z$ in $R_j^t$ belongs to $X_j$. 
We can assume without loss of generality that $c_j = (0,0)$ and $c_{j+1} = (0,v)$ with $v > 0$ by using a translation and a rotation. The set $X_j$ can then be rewritten as 
\[
X_j = \begin{cases}
    B^f( c_{j+1}, K) \, \cup \, \bigl( [C,D] + \mathbb{R}^-_y \bigr) \text{ if } j=0 \\
    B^f( c_j, K) \, \cup \, ABCD \, \cup \, B^f( c_{j+1}, K) \text{ if } j \in \llbracket 1, N \rrbracket
\end{cases}
\]
where $ABCD$ is the rectangle given by the points $A = (-K, 0)$, $B = (K, 0)$, $C = (K, v)$, $D = (-K, v)$ and $\mathbb{R}^-_y = \{ (0, \tau) \,: \, \tau \in \mathbb{R}^- \}$. It is clear that the set  $X_j$ is convex from this description. We represent in  \Cref{fig:lemme P_l} the sets $X_j$ with $X_0$ in (a), and the other cases in (b). 
    
\begin{figure}[h!]
\centering
  \begin{subfigure}[b]{.35\textwidth}
    \centering
      \includegraphics[
angle=-90,width=\textwidth]{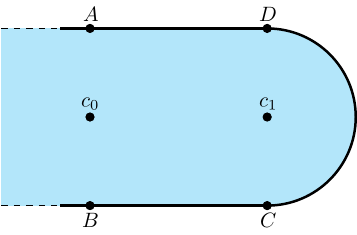}
      \caption{The set $X_0$}
  \end{subfigure}
  \centering
  \begin{subfigure}[b]{0.35\textwidth}
    \centering
     \includegraphics[
angle=-90,width=\textwidth]{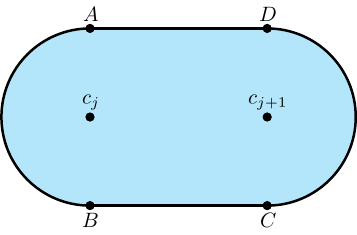}
      \caption{The set $X_j$ for $j = 1, \dots, N$}
  \end{subfigure}
  \caption{}
  \label{fig:lemme P_l}
\end{figure}

Let us consider the convex hull $E_j:= \text{Conv}(a_j, b_j ,a_{j+1}, b_{j+1})$, we obtain $R_j^t$ included in $E_j$. Now, as $a_j, b_j, a_{j+1}, b_{j+1} $ belong to $X_j$ (by result 1.) and that $X_j$ is convex, we then have $E_j \subseteq X_j$ and $R_j^t \subseteq X_j$.
\end{proof}

\begin{prop} \label{prop faisceau P_l condition non obscuration}
    Let $L$ be an $N$-reflexive polygonal chain given by $\bigl((c_i), (\theta_j)\bigr)$ and $F:= P_l(L)$ with  $l = \bigl( (l_j^a, l_j^b) \bigr)$, the primary beam given by $\bigl((a_i),(b_i)\bigr)$. Assume \[
    K:= \max\limits_{1 \leq j \leq N} \max \bigl( l_j^a, l_j^b \bigr) < \frac{d_O(L)}{2},
    \]
    then the \cref{condition non-obscuration} condition of a beam is verified by $F$. 
\end{prop}

\begin{proof}
    Let us take $w$ in $S_k = [a_k,b_k]$ and $z$ in $R_j^t$ with $0 \leq j \leq N$  and  $1 \leq k \leq N+1$ such that $k \neq j, j+1$. By  \cref{lemme faisceau P_l condition non obscuration}, we have 
    $|w-c_k| \leq K $ and $   |z-x| \leq K$
    with $x$ in the $j$-th ray $R_j$ of $L$. 
    Thus, we obtain
    \[
    |z-w| \geq |x-c_k| - |z-x| - |w-c_k| \geq d_O(L) - K - K
    \]
    since $|x-c_k| \geq d_O(L)$ by the definition of $d_O$. 
    So we have $ |z-w| \geq d_O(L) - 2K > 0$
    by the hypothesis $K < \frac{d_O(L)}{2}$ and then the non-obscuration condition $
    S_k \cap R_j^t = \varnothing
    $
    is verified.
\end{proof}

\begin{corollaire} \label{corollaire faisceau P_l}
    Let $L$ be an $N$-reflexive polygonal chain and $F$ given by $P_l(L)$. Assume 
    \[
    K:= \max\limits_{1 \leq j \leq N} \max (l_j^a,l_j^b) < \min\Bigl(d_R(L), \frac{d_O(L)}{2} \Bigr),
    \] 
    then $F$ is a beam. 
\end{corollaire}

\begin{proof}
    By \cref{prop faisceau P_l condition reflexion} and \cref{prop faisceau P_l condition non obscuration}, we have that $F$ verifies the two necessary conditions to be a beam.
\end{proof}

\begin{prop}[Homotopy with different parameters $\lambda $ and $\mu$] \label{prop homotopie a parametres differents}
    Let $L$ be an $N$-reflexive polygonal chain, $F:= P_\lambda (L)$ and $F':= P_{\mu}(L)$ be two beams given respectively by $((a_i),(b_i))$ and $((a_i'),(b_i'))$, with respective parameters $\lambda$ and $\mu$ such that 
    \[
    K,K' < \min\biggl(d_R(L), \frac{d_O(L)}{2}\biggr)
    \]
    with $K:= \max\limits_{1 \leq j \leq N} \max (\lambda_j^a,\lambda_j^b)$ and $K':= \max\limits_{1 \leq j \leq N} \max (\mu_j^a,\mu_j^b)$, then $F\cong F'$.
\end{prop}

\begin{proof}
Straightforward by defining the application $\gamma_{F,F'}:t\ni [0,1] \longmapsto (1-t)F + tF'\in \FaiscP$. By using that $\gamma_{F,F'}(t)=P_{(1-t)\lambda + t\mu}$ and \cref{corollaire faisceau P_l} we obtain that $\gamma_{F,F'}([0,1])\in \Faisc$ and eventually that $F$ and $F'$ are homotopic through $\gamma_{F,F'}$.
\end{proof}

\begin{defprop}[Subbeam]
\label{defprop subbeam}
    Let $F$ be an $N$-beam, we define the subbeam 
    $
    F_\mu:= \bigl( P_{t_a}, P_{t_b} \bigr)
    \in  \Faisc$
    where $t_a = \frac{1 - \mu}{2}$ and $t_b = \frac{1 + \mu}{2}$ with parameter $0 <  \mu \leq 1$ (see  \cref{def:t-ligne brisee}).
\end{defprop}

\begin{proof}
Let $F_\mu = \bigl((a_i^\mu),(b_i^\mu)\bigr)$, the reflexion condition is easily verified since $S_k^\mu = [a_k^\mu, b_k^\mu]\subset S_k= [a_k, b_k]$. 
Moreover, for all element $x \in R_{j,\mu}^t$ where $R_{j,\mu}^t$ is the $j$-th ray of the $t$-polygonal chain of $F_\mu$, there exists $\nu$ in $[0,1]$ such that $x$ belongs to $R_j^\nu$, with $R_j^\nu$ the $j$-th ray of the $\nu$-polygonal chain of $F$. 
So we have $R_{j,\mu}^t \, \cap \, S_k = \varnothing$ (with $0 \leq j \leq N$ and $1 \leq k \leq N+1$ such that $k \neq j,j+1$) by the non-obscuration condition of $F$. Thus, the non-obscuration condition for $F_\mu$ is verified what ends the proof.
\end{proof}

Let us notice that the beam $F$ can be written as $F_1$, the subbeam of $F$ with parameter $1$.

\begin{corollaire}[Homotopy of subbeams] \label{corollaire homotopie sous-faisceaux}
    Let $F$ be an $N$-beam and $F_\mu$ be a subbeam of $F$ with parameter $0 < \mu \leq 1$. We introduce the application 
    \begin{align*}
        H^F_\mu: [0,1] &\longrightarrow \Faisc \\
        t &\longmapsto F_{1 + t(\mu - 1)}
    \end{align*}
    connecting $F = F_1$ to $F_\mu$.
    So, $F$ and $F_\mu$ are homotopic.
\end{corollaire}

\begin{theo} \label{thm invariance}
    Let $F, F'$ be in $\Faisc$ such that $\IFaisc(F) = \IFaisc(F')$, that is to say $\NFaisc(F) = \NFaisc(F')$ and $\CarFaisc(F) = \CarFaisc(F')$, then $F \cong F'$, or equivalently, $F$ and $F'$ are in the same connected component of $\Faisc$.
\end{theo}

\begin{proof}
    In order to show the theorem, we first focus on reflexive polygonal chains produced by the beams $F$ and $F'$ with the application $M_R: \Faisc \rightarrow \LR$. That is why we write $L:= M_R(F)$ and $L':= M_R(F')$. As $\NR(L) = \NR(L')$ and $\Car(L) = \Car(L')$, we obtain by  \cref{prop_compo_LR} a path $\gamma_R: [0,1] \rightarrow \LR$ such that $\gamma_R (0) = L$ and $\gamma_R(1) = L'$.

    Let us come back now to the general case of beams. To rely on the case of reflexive polygonal chains, we will define an application $P^\varepsilon: B \rightarrow \FaiscP$ with parameter $\varepsilon > 0$ with $B = \text{Im}(\gamma_R) $. 
    We then define the application $P^\varepsilon: B \rightarrow \FaiscP$ as $P_{l_\varepsilon}$ (see  \cref{def application P_l}) with $l_\varepsilon = \bigl((\varepsilon, \dots, \varepsilon), (\varepsilon, \dots, \varepsilon)\bigr)$. Explicitly, $P^\varepsilon$ is defined as
    \begin{align*}
        P^\varepsilon: B &\longrightarrow \FaiscP \\
        ((c_i), (\theta_j)) &\longmapsto ((a_i),(b_i))
    \end{align*}
    with 
    \[
    (a_i, b_i) = \begin{cases}
        (c_i,c_i) \text{ if } i = 0, N+1, \\
        \bigl( c_i + \varepsilon (\cos(\theta_i),\sin(\theta_i)), c_i - \varepsilon ( \cos(\theta_i),\sin(\theta_i)) \bigr) \text{ if } i \in \llbracket 1,N \rrbracket
    \end{cases}
    \]
    We show there exists a parameter $\varepsilon>0$ small enough such that
    $
    \text{Im}(P^\varepsilon) \subseteq \Faisc$.
    We must then check the two conditions for a primary beam to be a beam. In the following steps, we verify one after the other the reflexion and non-obscuration conditions of a beam:
    \begin{enumerate}
        \item[$\circ$] (Reflexion) \label{etape 1 epsilon}  
        Recall that we have $\gamma_R: [0,1] \rightarrow \LR$, the path between $L = M_R(F)$ and $L' = M_R(F')$. We now define
        $
            \widetilde{d}_R = d_R \circ \gamma_R: [0,1] \longrightarrow \mathbb{R}^+_* 
        $
        that is continuous. As this function is continuous on a compact domain, the minimun of the function is reached and we write
        $
        m_R:= \min\limits_{t \in [0,1]} \widetilde{d}_R(t) > 0$.
        By taking $\varepsilon < m_R$, we have $P^\varepsilon(b)$ that verifies the reflexion condtion of beams for all $b$ in $B$ by  \cref{prop faisceau P_l condition reflexion}. 
        
        \item[$\circ$] (Non-obscuration) \label{etape 2 epsilon} 
        Let us now consider $      
            \widetilde{d}_O = d_O \circ \gamma_R: [0,1] \longrightarrow \mathbb{R}^+_* 
        $
       and notice that $\widetilde{d}_O$ is continuous on a compact. Thus, the minimun of the function is reached and we write
        $
            m_O:= \min\limits_{t \in [0,1]} \widetilde{d}_O(t) > 0
        $.
        By taking $\varepsilon < \frac{m_O}{2}$, we have $P^\varepsilon(b)$ that verifies the non-obscuration condition of beams for all $b$ in $B$ by  \cref{prop faisceau P_l condition non obscuration}. 
    \end{enumerate}
With the two steps above, we take 
    $\varepsilon < \min \Bigl( m_R , \frac{m_O}{2} \Bigr)$
and we then have $P^\varepsilon(\gamma_R(t))$ in $\Faisc$ for all $t$ in $[0,1]$. 
With the condition on $\varepsilon$, we obtain the function  $
P_{F^\varepsilon}^{(F')^\varepsilon}:= P^\varepsilon \circ \gamma_R: [0,1] \longrightarrow \Faisc$
connecting $F^\varepsilon:= P^\varepsilon(F)$ to $(F')^\varepsilon:= P^\varepsilon(F')$. \\
Now we must show
$ F \cong F^\varepsilon$ and $F' \cong (F')^\varepsilon $
in order to get a beam homotopy between $F$ and $F'$. 
We will show that $F$ is homotopic to $F^\varepsilon$ in two steps. In step \ref{homotopie etape 1}, we show that $F$ is homotopic to $F_\mu$, subbeam of $F$ with a good choice of $\mu$ for step \ref{homotopie etape 2} where we show that $F_\mu$ and $F^\varepsilon$ are homotopic. 
\begin{enumerate}
    \item \label{homotopie etape 1} 
    We write $\bigl((a_i^\mu), (b_i^\mu)\bigr)$ for the subbeam $F_\mu$ of $F$. It can be rewritten in the form of $P_l(L)$ (we recall that $L = M_R(F)$) with $l_\mu = \Bigl(\bigl((l_\mu)_j^a \bigr)_{1 \leq j \leq N} , \bigl( (l_\mu)_j^b \bigr)_{1 \leq j \leq N} \Bigr)$ which is given explicitly by 
    $
    (l_\mu)_j^a = \Bigl| P_j^\frac{1}{2} - a_j^\mu \Bigr| $ and $(l_\mu)_j^b = \Bigl| P_j^\frac{1}{2} - b_j^\mu \Bigr|
    $
    with $j$ in $\llbracket 1,N \rrbracket$ where $\bigl(P^\frac{1}{2}_i\bigr)$ is the centered polygonal chain of $F$ (see  \cref{def:ligne brisee centrale}). \\
    In the next step, we will take $\mu$ in $]0,1]$ low enough to get 
    \[
    K_\mu:= \max\limits_{1 \leq j \leq N} \max \bigl((l_\mu)_j^a,(l_\mu)_j^b \bigr) < \min\Bigl(d_R(L), \frac{d_O(L)}{2} \Bigr).
    \]
    We consider now the $H_\mu^F$  from  \cref{corollaire homotopie sous-faisceaux} connecting $F$ to $F_\mu$.
    \item \label{homotopie etape 2}
    Thanks to the condition on $\mu$ considered in the previous step, we have 
    $
    K_\mu < \min \bigl(d_R(L),\frac{d_O(L)}{2} \bigr) 
    $
    and we also have
    $
    \varepsilon < \min \bigl(d_R(L),\frac{d_O(L)}{2} \bigr),
    $
    so we obtain by  \cref{prop homotopie a parametres differents} the path $\gamma_{F_\mu, F^\varepsilon}$ between $F_\mu$ and $F^\varepsilon$.
\end{enumerate}
Thus, we consider the concatenation of paths $P_F:= H^F_\mu \cdot \gamma_{F_\mu, F^\varepsilon}$ (see footnote \ref{footnote:concat} for the concatenation of paths) and we have $P_F$ which connects $F$ to $F^\varepsilon$ and so $F \cong F^\varepsilon$. \\
The same reasoning can be done with $(F')^\varepsilon$ and $F'$ by considering $F'_\nu$ with an adapted $\nu$. Thus, we obtain in the same manner a path $P^{F'}$ connecting $(F')^\varepsilon$ to $F'$. Finally,  the path
$
P_{F}^{F'}:= (P_F \cdot P_{F^\varepsilon}^{(F')^\varepsilon}) \cdot P^{F'}  $
connects the beam $F$ to the beam $F'$.
\end{proof}

\begin{corollaire}
    We have a characterization of the connected component of $\Faisc$, we have: 
    \[
    \pi_0 \bigl( \Faisc \bigr) \simeq \pi_0 \bigl( \LO \bigr) \times \{ -1, 1 \}^N
    \]
    with the application
    \begin{align*}
        \IFaisc: \Faisc &\longrightarrow \pi_0(\LO) \times \{ -1, 1 \}^N  \\
        F &\longmapsto \bigl(\NFaisc(F), \CarFaisc(F)\bigr)
    \end{align*}
\end{corollaire}

\begin{proof}
Let $F,F'$ be beams in $\Faisc$. If $F$ and $F'$ are in the same connected component, then we obtain $\IFaisc(F) = \IFaisc(F')$ by  \cref{prop:invariance}. \\
Reciprocally, if we have $\IFaisc(F) = \IFaisc(F')$ then $F$ and $F'$ are in the same connected component by  \cref{thm invariance}. \\
So, we find 
\[ 
\pi_0 \bigl(\Faisc \bigr) \simeq \pi_0 \bigl( \LO \bigr) \times \{ -1, 1 \}^N
\]
by noticing that $\IFaisc$ is a surjective application.
\end{proof}
We have shown that $\IFaisc$ is an exact topological invariant according to  \cref{def:invariant}.

\subsection{Symmetrical case} \label{subsection:FaiscSym}
In  \cref{subsection:Faisc}, we have defined a beam $F$ as the data of two polygonal chains $\bigl( (a_i),(b_i) \bigr)$. If we consider now the beam $F'$ defined as $\bigl( (b_i), (a_i))$, we obtain two different beams which have nevertheless the same geometry when we represent them: the order does not have an influence on the interpretation of the beam. That is why we introduce in this subsection the space of symmetrical beams $\FaiscS$; the beams $F$ and $F'$ then produce the same symmetrical beam. 
\begin{defi}[Symetrical beams]
    The discrete group $G = \mathbb{Z} / 2 \mathbb{Z}$ acts continuously on the space of beams $\Faisc$ by taking the action 
    \begin{align*}
        0 \cdot \bigl((a_i),(b_i)\bigr) &= \bigl((a_i),(b_i)\bigr) \\
        1 \cdot \bigl((a_i),(b_i)\bigr) &= \bigl((b_i),(a_i)\bigr)
    \end{align*}
    where $\mathbb{Z}/ 2\mathbb{Z}$ has two elements, $0$ (identity element) and $1$, with $1 + 1 = 0$ and where $\bigl((a_i),(b_i)\bigr)$ is a beam. \\
    We then define the space of \emph{symmetrical beams} $\FaiscS$ as 
    \[
    \FaiscS = G \backslash \Faisc .
    \]
\end{defi}
The orbit $O_F$ of an $N$-beam $F$ given by $\bigl((a_i),(b_i)\bigr)$ with the action of $G$ has two elements
\[
O_F = \{ F, \FI \}
\]
with $\FI = \bigl((b_i),(a_i)\bigr)$ that we call the inverse beam of $F$. The inverse beam is indeed different from $F$ since $a_1 \neq b_1$ (condition on the beams).
Let us introduce the signature of a beam in $\FaiscS$:
\begin{defi}[Signature] \label{def:signature faisceau}
    Let us define the \textit{signature} application as
    \begin{align*}
        \sign^F: \Faisc &\longrightarrow \{ -1, 1 \}^{N-1} \\
        F &\longmapsto \sign^F(F)
    \end{align*}
    with
    \[
    \sign^F(F)_k = \begin{cases}
        1 \text{ if } \CarFaisc(F)_k \text{ and } \CarFaisc(F)_{k+1} \text{ have the same sign,} \\
        -1 \text{ otherwise} \end{cases}
    \]
    for $1 \leq k \leq N-1$.
\end{defi}

We then define a topological invariant $\IFaiscSym:= (\NFaiscSym \times \signSym)$ for $\FaiscS$ which is the conjunction of the two following invariants:
\begin{itemize}
    \item[$\circ$] the $\NFaiscSym$ application that comes from $\NFaisc$,
    \item[$\circ$] the $\signSym$ application that comes from $\sign^F$ (see \cref{def:signature faisceau}). We cannot indeed use the characteristic as an invariant since we have $\CarFaisc(F) = -\CarFaisc(F')$ but we have $\sign^F(F) = \sign^F(F')$,
\end{itemize}
through the quotient application $\pi_S: \Faisc \rightarrow \FaiscS$.
Then it is straightforward to show that the invariant $\IFaiscSym$ is exact with 
\begin{theo} \label{thm invariance symetrique}
Let $\mathcal{F}$, $\mathcal{F}'$ be in $\FaiscS$ such that $\IFaiscSym(\mathcal{F}) = \IFaiscSym(\mathcal{F}')$, which means $\NFaiscSym(\mathcal{F}) = \NFaiscSym(\mathcal{F'})$ and $\signSym(\mathcal{F}) = \signSym(\mathcal{F}')$, then $\mathcal{F}$ and $\mathcal{F}'$ are in the same connected component of $\FaiscS$. 
\end{theo}

\section{Application} \label{section:application}
It has been shown that the connected components of the set of non-obscuring solutions that satisfy a reflection condition  are characterized by the invariant outlined in \cref{thm invariance} and \cref{thm invariance symetrique}.  By comparing this invariant with that presented in \cite{onaxis24}, which is related to the set of three-mirror telescope solutions defined by polynomial equations corresponding to first-order optical conditions, it becomes evident that the orientation of the mirrors is a shared characteristic.  As noted in \cite{onaxis24}, by incorporating information about the signs of the curvatures, we derive a new invariant for this paired model that we refer to as \textit{on-off}. It is worth noticing that this invariant is quite practical for optical designers in classifying first-order obscuration-free solutions during the exploratory process. 

Let us clarify the connection with \cite{onaxis24} in more detail. We define the \emph{magnifications} 
$(\Omega_k)_{1 \leq k \leq N-1}$ of a beam \( F \) as follows:
\begin{equation}
\Omega_k = \sign(F)_k \times \frac{|b_{k+1} - a_{k+1}|}{|b_k - a_k |}.
\label{def magnifications}
\end{equation}
The magnifications, together with the entrance pupil diameter (which corresponds to the data of \( |a_1-b_1| \)), completely determine the polygonal chain \( (b_i) \) such that the condition in \eqref{def magnifications} holds true. Furthermore, from \cref{corollaire faisceau P_l}, we know that there exists a subbeam \( F_\mu \) with \( 0<\mu<1 \) that is sufficiently small such that \( F_\mu \) qualifies as a beam (see \cref{defprop subbeam}).
Hence, let us examine a polygonal chain \( L := (a_i) \). We can compute the signed distances, defined as \( d_k = (-1)^k |a_{k+1} - a_k| \) for \( k=1,\dots,N \). Following the approach in \cite{onaxis24}, for \( N=3 \) mirrors in the codimensional 2 focal case, we consider the set 
\[
E_{on} = \{ x = (d_1, d_2, d_3, \Omega_1, \Omega_2) \in \mathbb{R}^5 \mid (f_1(x) = 0) \land (f_2(x) = 0) \land \mathcal{C}(x) \},
\]
composed of the real solutions to the two first-order polynomial equations known as the \emph{focal} equation (\( f_1 = 0 \)) and the \emph{Petzval} equation (\( f_2 = 0 \)), subject to a non-degeneracy constraint \( \mathcal{C}(x) = \land_{k=1}^3 [(-1)^k d_k > 0) \land (c_k(x) \neq 0)] \) where $c_k(x)$ is the  curvature of the $k$-th mirror.
 It is shown in \cite{onaxis24} that  for a given sequence of distances \( (d_1, d_2, d_3) \) and a focal value $f\in \{-1,1\}$, there exist two solutions if and only if a discriminant, which depends on these distances, is positive.
Therefore, let \( (d_1, d_2, d_3) \) and a focal \( f \in \{-1, 1\} \). If the discriminant derived from \( L \) is positive or zero, we obtain two solutions in \( (\Omega_1, \Omega_2) \) along with an on-axis nomenclature that encodes the signs of the magnifications corresponding to \( \signSym \) and the curvatures of the mirrors.
For instance, the designation \({PP101}\) signifies that the first and second magnifications are positive, the first mirror is convex, the second mirror is concave, and the last mirror is convex. In optical terminology, a configuration referred to as \({PP101}\) (or more accurately its anastigmat version, which is an optical configuration that also eliminates third-order aberrations by selecting appropriate conicity values) corresponds to a Korsch telescope. This type of telescope is frequently utilized in space observation missions (e.g., Euclid, James Webb, etc.).

\begin{figure}[h!]
    \centering
     \includegraphics[width=.7\linewidth]{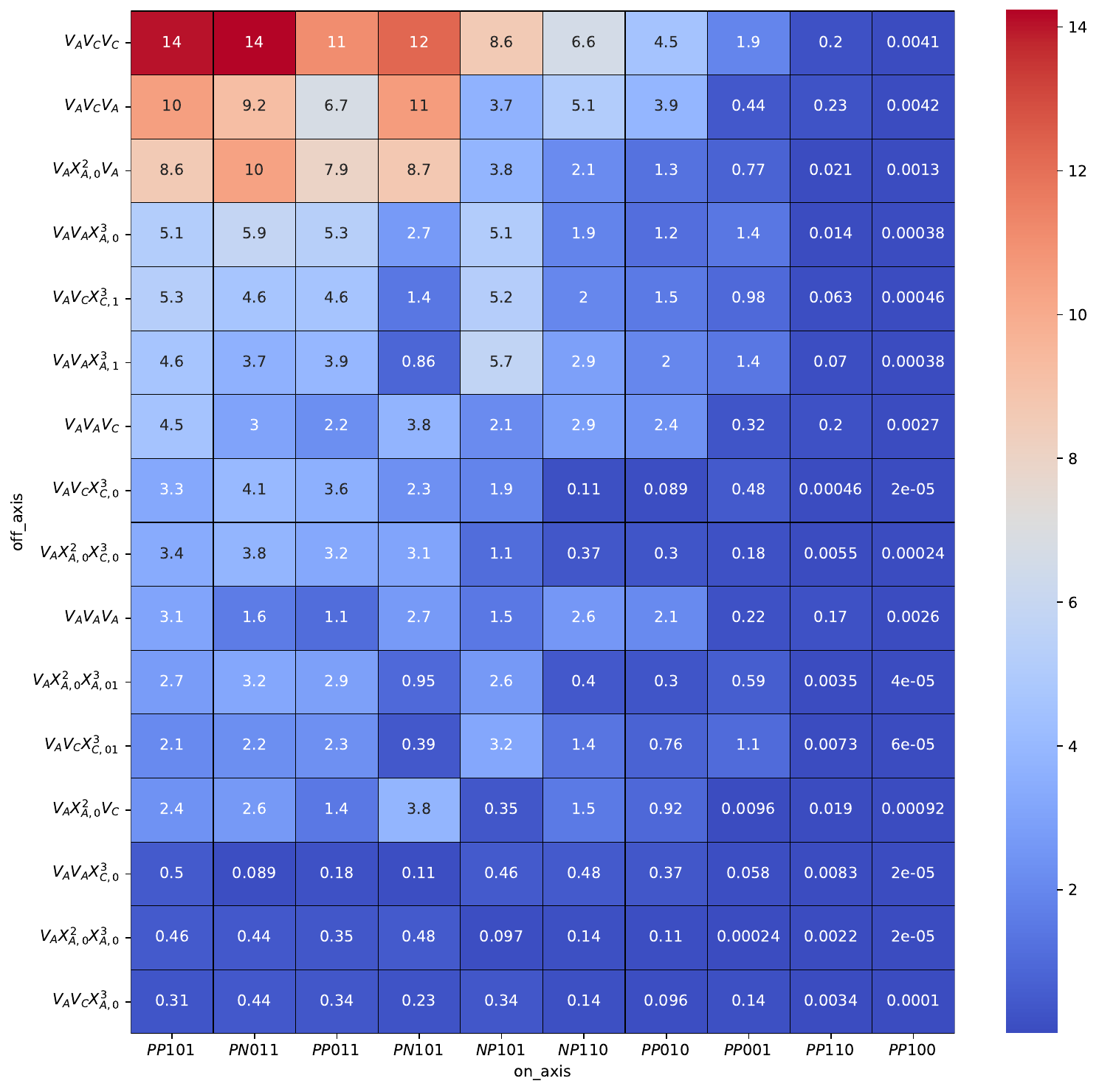}
    \caption{\textit{On-off} axis topological invariants repartition estimation for $N=3$ mirrors in the space of polygonal chains contained in an hyper rectangle of length $2.6\times\text{focal}$.}
    \label{fig:classification off-on axis}
\end{figure}

\begin{figure}[h!]
    \centering
    \begin{subfigure}[b]{0.32\textwidth}
        \centering
        \includegraphics[trim=3cm 3cm 3cm 3cm, clip, width=\textwidth]{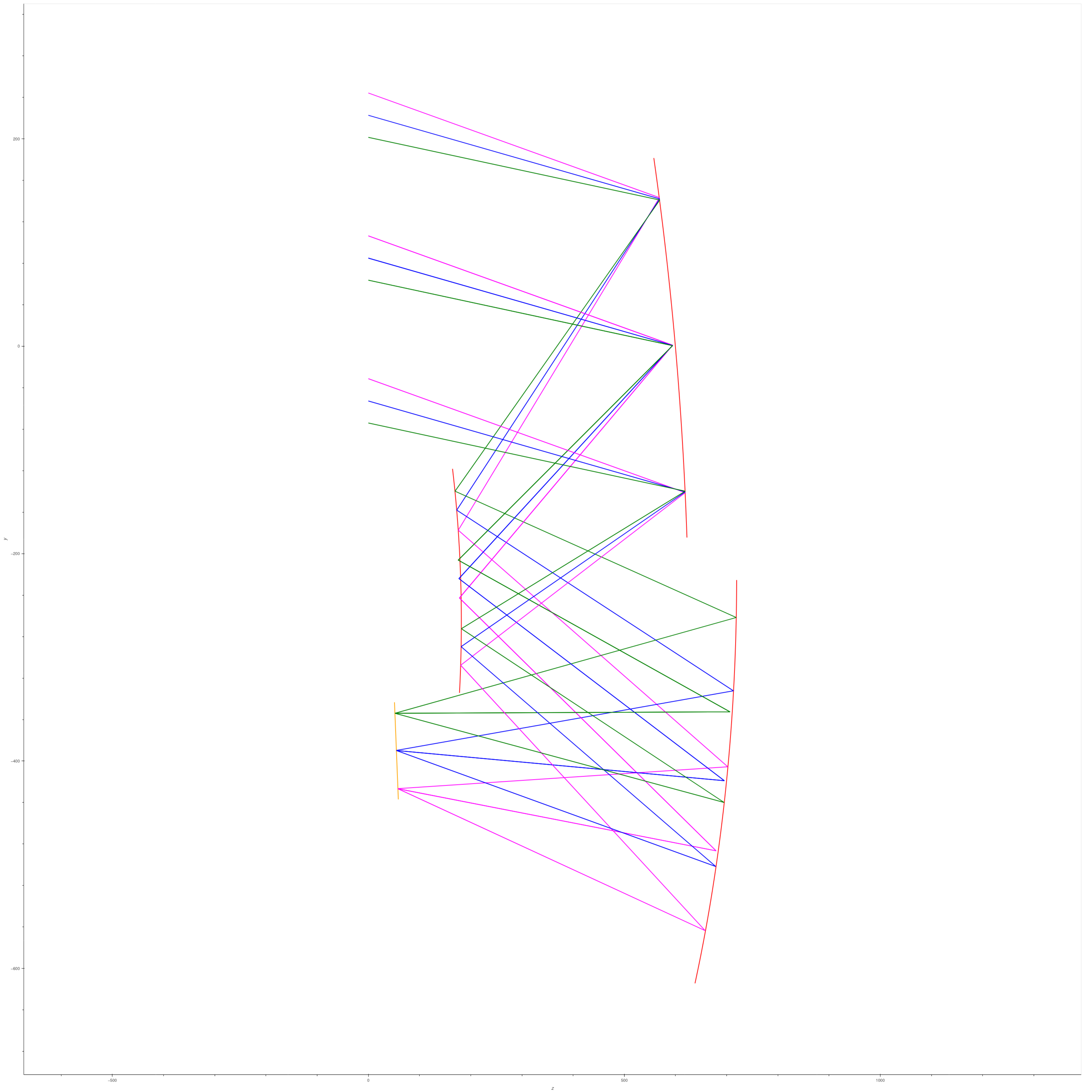} 
        \caption{$PP101-V_AV_CV_A$}
        \label{fig:sub1}
    \end{subfigure}
    \begin{subfigure}[b]{0.32\textwidth}
        \centering
        \includegraphics[trim=3cm 3cm 3cm 3cm, clip, width=\textwidth]{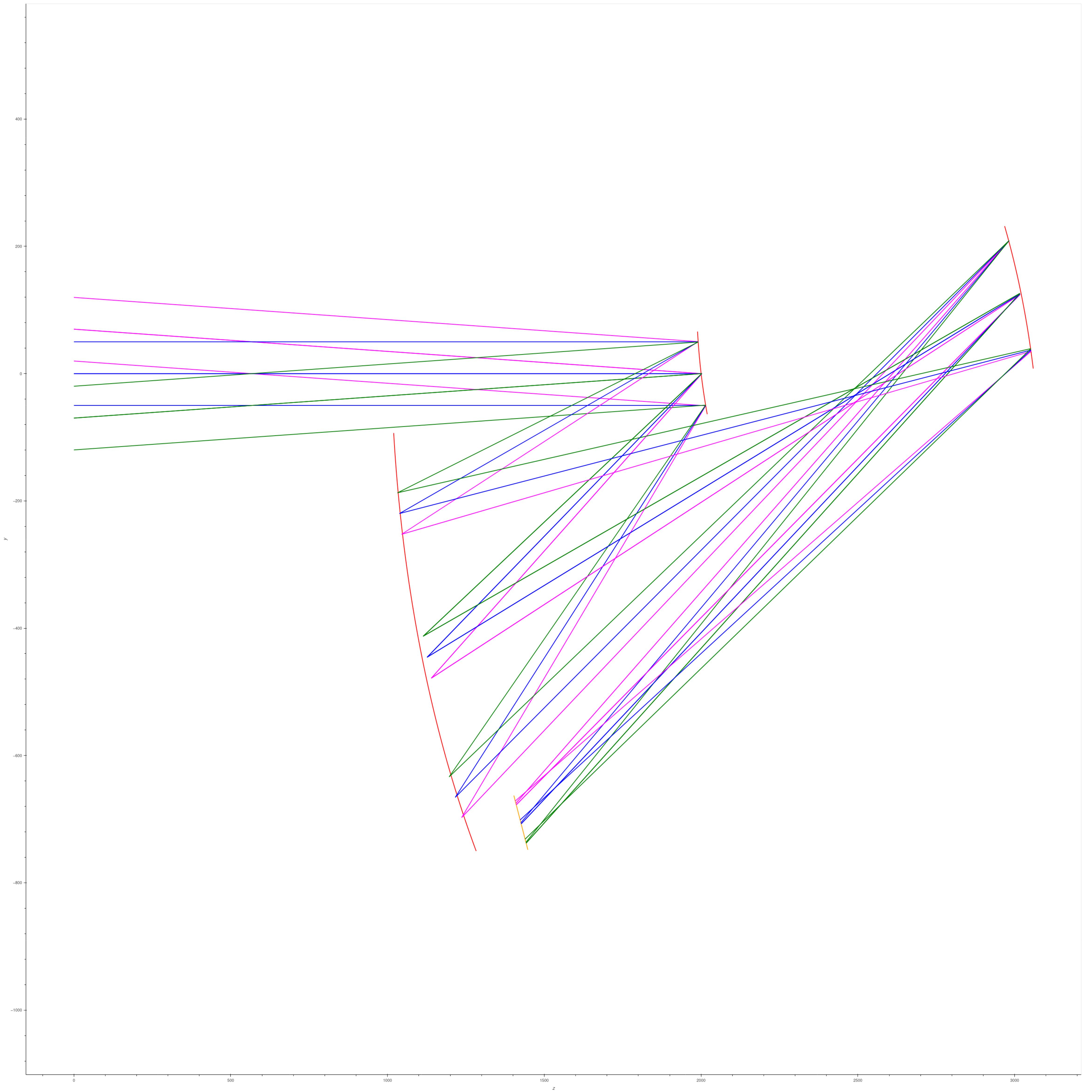} 
        \caption{$PN011-V_AV_CV_A$}
        \label{fig:sub2}
    \end{subfigure}
    \begin{subfigure}[b]{0.32\textwidth}
        \centering
        \includegraphics[trim=3cm 3cm 3cm 3cm, clip, width=\textwidth]{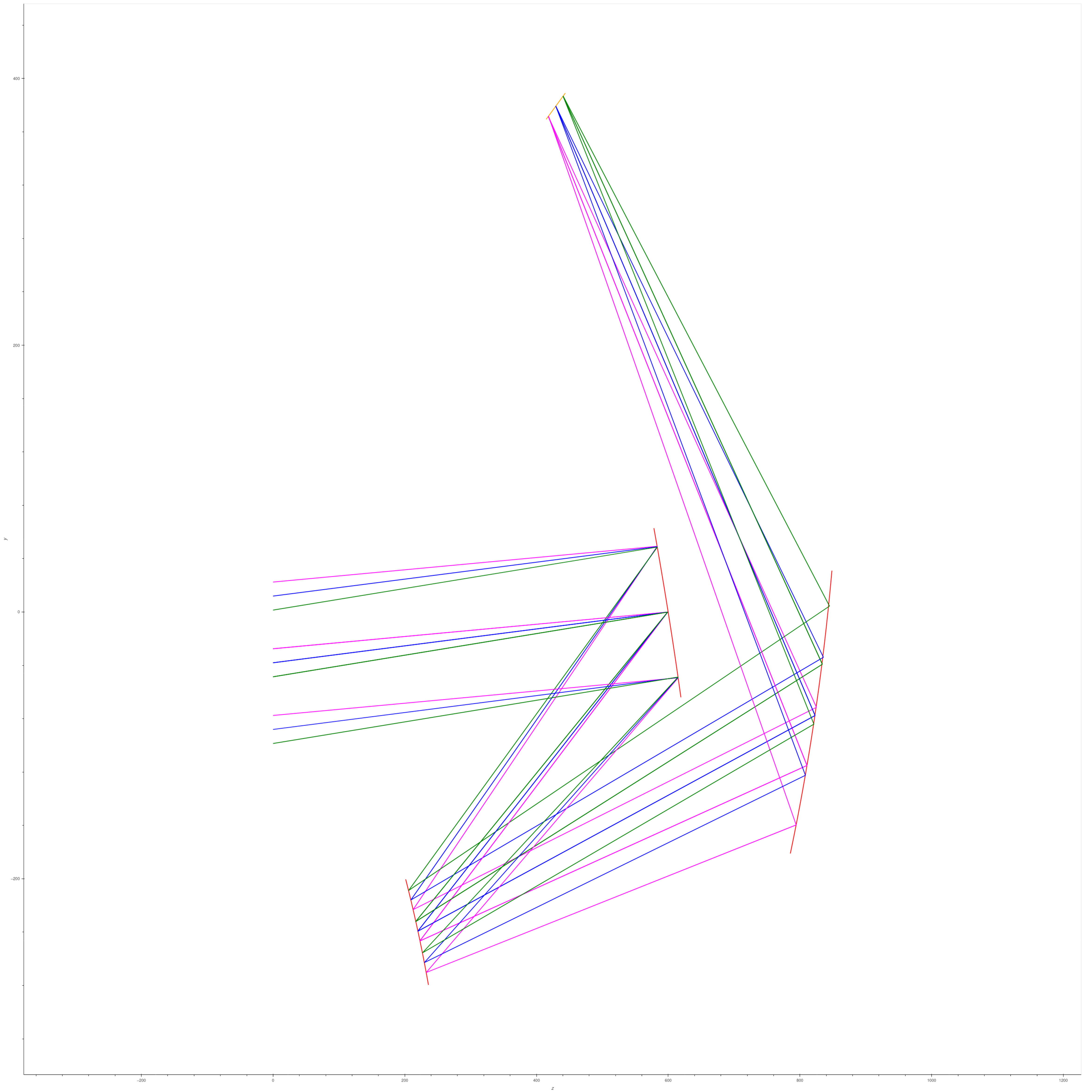} 
        \caption{$PP101-V_AV_CV_C$}
        \label{fig:sub3}
    \end{subfigure}
    \hfill
    \vspace{1em} 
    \begin{subfigure}[b]{0.32\textwidth}
        \centering
        \includegraphics[trim=3cm 3cm 3cm 3cm, clip, width=\textwidth]{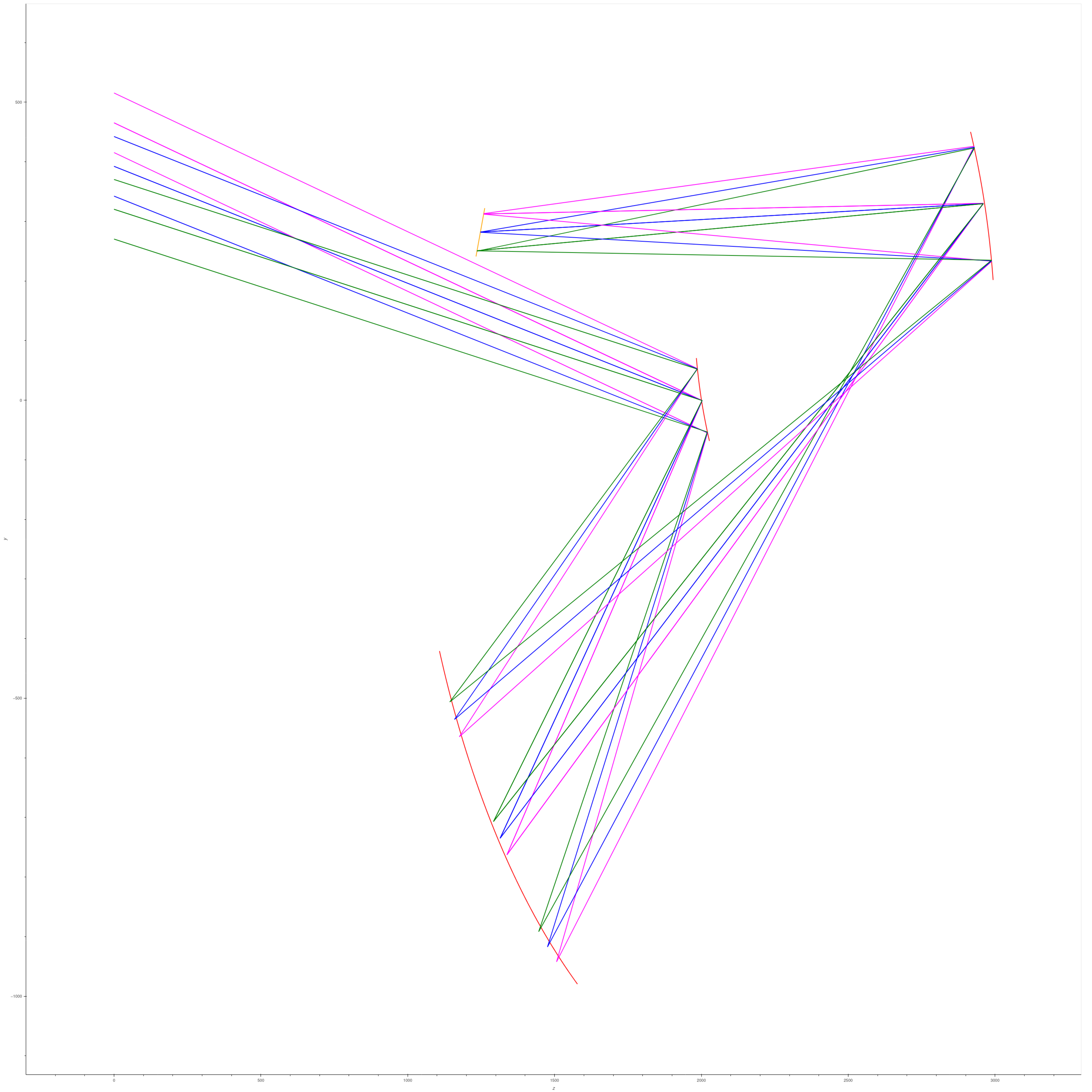} 
        \caption{$PN011-V_AV_CV_C$}
        \label{fig:sub4}
    \end{subfigure}
     \begin{subfigure}[b]{0.32\textwidth}
        \centering
        \includegraphics[trim=3cm 3cm 3cm 3cm, clip, width=\textwidth]{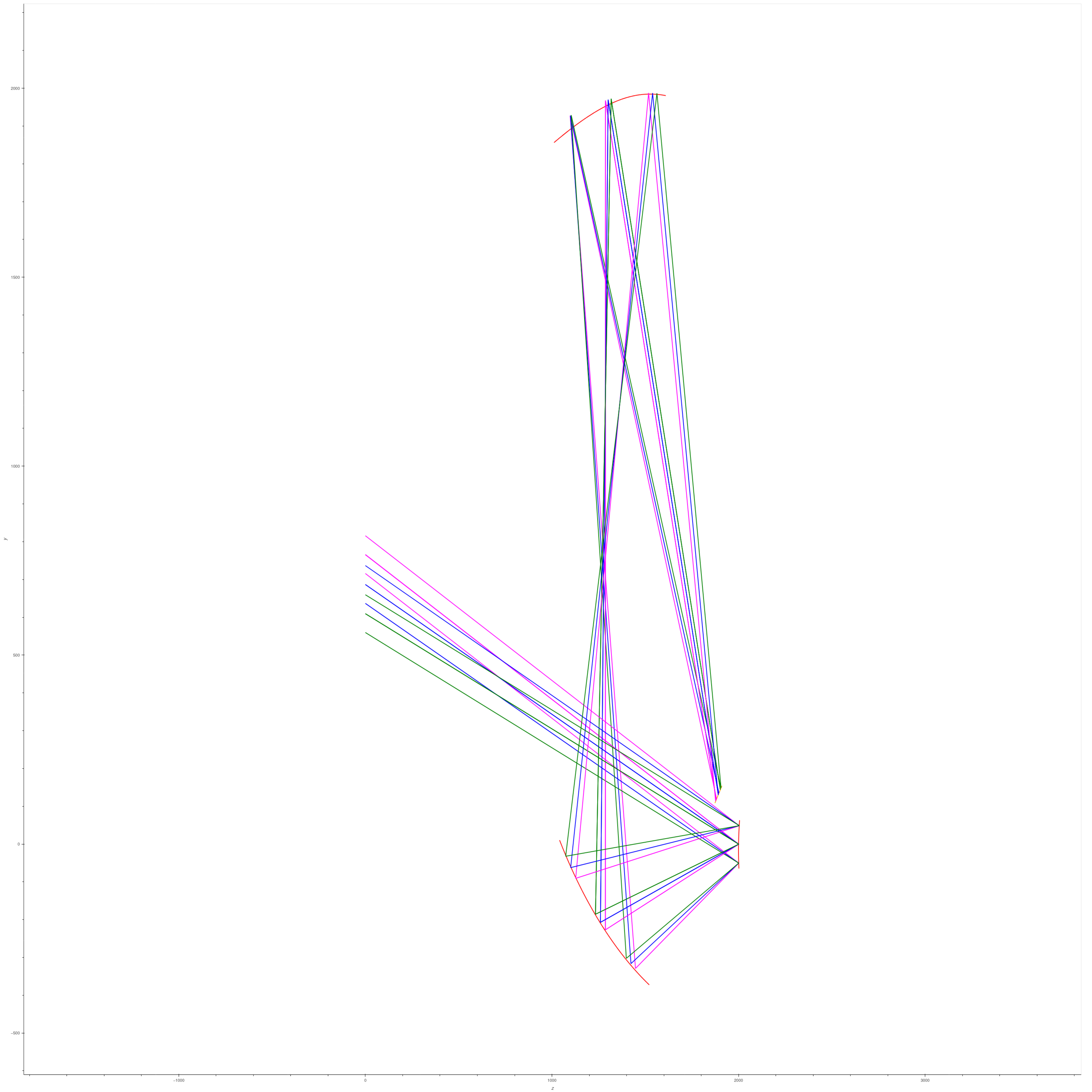} 
        \caption{$PN011-V_AX_{A,0}V_A$}
        \label{fig:sub4}
    \end{subfigure}
\hfill
    \caption{Different classes of three mirrors obscuration-free first-order controlled telescopes.}
    \label{fig:examples}
\end{figure}

To numerically illustrate the distribution of paired \emph{on-off} axis topological invariants for three-mirror focal obscuration-free telescopes (that are in \( E_{on} \) and in the space of obscuration-free polygonal chains), we sample \( N \) polygonal chains \( (a_i) \in \text{LO}^3(\mathbb{R}^2) \) from a uniform distribution. This sampling verifies the following conditions: \( a_0 = (0, 0) \), \( a_1 = (0, r) \), \( a_2 \) is chosen from the range \( [0, r] \times [-r, r] \), and both \( a_3 \) and \( a_4 \) are selected from the square defined by \( [-r, r]^2 \) for \( r > 0 \).
According to \cite{onaxis24}, all projections of the connected components of \( E_{on} \) onto the distance space \( (d_1, d_2, d_3) \) intersect the hyper-rectangle \([-1, 0] \times [0, 1] \times [-1, 0]\). Therefore, we set \( r = 1.3 \) and $N=3\times 10^6$ to ensure that there is a non-zero probability of obtaining at least one point in each component of \( E_{on}  \). In \Cref{fig:classification off-on axis}, we illustrate the distribution of optical systems based on their on-off axis nomenclature within the space defined by \( (a_2, a_3, a_4) \in ([0, r] \times [-r, r]) \times [-r, r]^2 \times [-r, r]^2 \).
For example, for a given focal $f\in \{-1,1\}$,  there is $14$ percent of chance to get the  configuration ${PP101-V_AV_CV_C}$ and the configuration ${PN011-V_AV_CV_C}$ while there is only $4.10^{-5}$ percent of chance to get a ${PP100-V_AX_{A,0}^2X_{A,01}^3}$ configuration.
This numerical examples show that there are at least 160 connected components in the paired space. 
\Cref{fig:examples} illustrates one represent  of 5 different  on-off axis classes. For details on the names we refer to \cite{onaxis24,offaxis24}.
While we cannot definitively answer whether this new invariant is exact for the paired model, the work presented here is a step further in that direction. 

\section*{Conclusion and prospects}
In this paper, we have extended the investigation of the connected components of the set of polygonal chains $\LO$, conducted in a previous work \cite{offaxis24}, to the connected components of the set of polygonal beams $\Faisc$.  Specifically, we demonstrate that an exact topological invariant can be obtained by incorporating the orientations of the mirrors alongside the one used to characterize the connected components of the set of polygonal chains. We finally deduce an exact topological invariant for the quotient space $\FaiscS$ obtained by the left action of the group $\mathbb{Z}/2\mathbb{Z}$ on $\Faisc$. By combining this with the signs of the curvatures of the mirrors, we illustrate the distribution of the various topological names obtained in the context of three-mirror focal telescopes constrained by first-order optical polynomial equations \cite{onaxis24}. For this paired model, we conjecture that there are 160 connected components. Future work will focus on providing a mathematical proof for this statement.
\bibliographystyle{my_plain}

\end{document}